\documentclass[prl,singlecolumn,a4paper,amsmath,amssymb]{revtex4}

\usepackage{geometry}
\usepackage{amsfonts}
\usepackage{latexsym}
\usepackage{graphicx}
\usepackage[usenames]{color}
\usepackage[utf8]{inputenc}
\usepackage[english,bulgarian]{babel}
\usepackage{hyperref}
\usepackage{caption}
\usepackage{morefloats}
\usepackage{epstopdf}
\usepackage{systeme}
\usepackage{listings}
\usepackage{array}
\usepackage{circuitikz}
\usepackage{tikz}

\newcommand{\be}{\begin{equation}}
\newcommand{\ee}{\end{equation}}
\newcommand{\nn}{\nonumber}

\begin{document}

\title{Measurement of the Boltzmann constant by Einstein. Problem of the 5-th Experimental
Physics Olympiad. Sofia 9 December 2017}


\author{Todor~M.~Mishonov, Emil~G.~Petkov, Aleksander~A.~Stefanov, Aleksander~P.~Petkov}
\email[E-mail: ]{mishonov@gmail.com, mishonov@bgphysics.eu}
\affiliation{Physics Faculty,\\
St. Clement of Ohrid University at Sofia,\\
5 James Bourchier blvd, BG-1164 Sofia}
\author{Iglika~M.~Dimitrova}
\affiliation{
Faculty of Chemical Technologies, Department of Physical Chemistry,\\
University of Chemical Technology and Metallurgy,\\
8, Kliment Ohridski blvd, BG-1756 Sofia}
\author{Stojan~G.~Manolev}
\email[E-mail: ]{manolest@yahoo.com}
\affiliation{Middle school Goce Delchev,\\
Purvomaiska str.~3, MKD-2460 Valandovo, R.~Macedonia}
\author{Simona~I.~Ilieva}
\email[E-mail: ]{simonailieva24@gmail.com}
\affiliation{Department of Atomic Physics, Physics Faculty,\\
St. Clement of Ohrid University at Sofia,\\
5 James Bourchier blvd, BG-1164 Sofia}
\author{Albert~M.~Varonov}
\email[E-mail: ]{akofmg@gmail.com}
\affiliation{Department of Theoretical Physics, Physics Faculty,\\
St. Clement of Ohrid University at Sofia,\\
5 James Bourchier blvd, BG-1164 Sofia}

\date{\today}

\begin{abstract}
Several consecutive experiments with specifically built set-up are described.
Performing of the consecutive experimental tasks enables possibility to determine Boltzmann's constant 
$k_\mathrm{_B}$. 
The fluctuations of the voltage $U(t)$ of series of capacitors connected in parallel
with a constant resistance are measured.
The voltage is amplified 1~million times $Y=10^6$. 
The amplified voltage $YU(t)$ is applied to a device, which give
the voltage mean squared in time
$U_2=\left<(Y U(t))^2\right>/U_0$.
This voltage $U_2$ is measured with a multimeter.
A series of measurements gives the possibility to determine the Boltzmann's constant
from the equipartition theorem $C\left<U^2\right>=k_\mathrm{_B}T$.
In order to determine the set-up constant $U_0$
a series of problems connected with Ohm's law are given that are addressed to the
senior students.
For the junior high school students, the basic problem is to analyse the analog mean squaring.
The students works are graded in four age groups S, M, L, XL.
The last age group contains problems that are for university students (XL category) and include
theoretical research of the set-up as an engineering device.
This problem is given at the Fifth Experimental Physics Olympiad ``Day of the Electron'',
on December 2017 in Sofia, organized by the Sofia Branch of the Union of Physicists in Bulgaria with the cooperation of the Physics Faculty of Sofia University and
the Society of Physicists of the Republic of Macedonia, Strumica. 
\end{abstract}
\captionsetup{labelfont={normalsize},textfont={small},justification=centerlast}
\maketitle

\section{Introduction}

From its very beginning, the Experimental Physics Olympiad (EPO) is worldwide known;
all Olympiad problems have been published in Internet~\cite{EPO1,EPO2,EPO3,EPO4} and from the very beginning there were 120 participants.
In the last years high-school students from 7 countries participated and the distance between the most distant cities is more than 4~Mm.

Let us describe the main differences between EPO and other similar competitions.
\begin{itemize}
\item Each participant in EPO receives as a gift from the organisers the set-up, which one worked with.
So, after the Olympiad has finished, even bad performed participant is able to repeat the experiment and reach the level of the champion.
In this way, the Olympiad directly affects the teaching level in the whole world.
After the end of the school year, the set-up remains in the school, where the participant has studied.
\item Each of the problems is original and is connected to fundamental physics or the the understanding of the operation of a technical patent.
\item The Olympic idea is realised in EPO in its initial from and everyone willing to participate from around the world can do that.
There is no limit in the participants number.
On the other hand, the similarity with other Olympiads is that the problems are direct illustration of the study material and alongside with other similar competitions mitigates the secondary education degradation, which is a world tendency.
\item One and the same experimental set-up is given to all participants but the tasks are different for the different age groups, the same as the swimming pool water is equally wet for all age groups in a swimming competition.
\end{itemize}

We will briefly mention the problems of former 4 EPOs. 
The setup of EPO1 was actually a student version of the American patent for auto-zero
and chopper stabilised direct current amplifiers~\cite{EPO1}.
The problem of the second EPO~\cite{EPO2} was to measure Planck constant by diffraction of a LED light by a compact disk.
A contemporary realization of the assigned to NASA patent for the use of negative impedance converter 
for generation of voltage oscillations was the set-up of EPO3~\cite{EPO3}.
EPO4~\cite{EPO4} was devoted to the fundamental physics -- to determine the speed of light by
measuring electric and magnetic forces.
The innovative element was the application of the catastrophe theory in the analysis of the stability of a pendulum.
In short, the established traditions is a balance between contemporary working technical inventions and fundamental physics.

And this 5$^\mathrm{th}$ Olympiad (EPO5) follows the established tradition.
In all countries study programs it is mentioned that heat is connected with atomic motion.
But even a comparatively simple example of the Maxwellian velocity distribution of molecules is not illustrated with an experiment even in very good universities.
The reason for this is because vacuum technology is very expensive and requires professional work.
On the other hand, the electronic measurements are foolproof and thousand times cheaper.
That is why, if we want to study the energy of the thermal fluctuations that is described by the temperature and Boltzmann constant~\cite{McCombie:71}
\be
m \left< v_x^2 \right>= k_\mathrm{_B} T= C \left <U^2 \right >,
\nn
\ee
the study of the mean square of the voltage $\left <U^2 \right >$ is thousand times cheaper than the measurement of the mean square of the velocity $\left< v_x^2 \right>$.
That is why, for methodological purposes for known capacity, the Boltzmann constant determination via electric measurements is the only possible method for high-school education.
As fundamental physics, this idea was proposed by Albert Einstein and now, 111 years later, the method is an experimental problem for high school students and the set-up has been produced in 200 copies.

\section{Current EPO5 problem}
111 years ago Albert Einstein~\cite{Einstein:07}offered a method for Boltzmann's
constant measurement $k_\mathrm{_B}$ by examination of the mean thermal energy
$C\left<U^2\right>=k_\mathrm{_B} T$.
Strangely, that experiment has never been conducted so far and you are the first ones
that will be able to do it after our team makes this idea a reality.
The technological progress during the last century made possible
that scientific problem to be adapted to the level of a school problem.
We begin with the tasks that can be solved by the junior students and
the difficulty arises gradually.
Each participant works as far as he/she can.
The last tasks are for the university students.

Tasks for further work.
The participants in the Olympiad receive from the organizers the experimental set-up 
they have worked with, as one of the most important goals of the Olympiad is the
careful repetition of the experiment up to the point of having a detailed measurement of the Boltzmann's
constant and the comprehensive understanding of the theory.
The problem's solution will be published on the Cornell University library server and
some details of scanned works of participants will be published
on the Sofia Branch of Union of Physicists in Bulgaria website.
After reading the solution,
we expect the participants to conduct all described experiments in it
and obtain the Boltzmann's constant.
$k_\mathrm{_B}$. 
This first determination of a fundamental constant throws upon us
the glare of a great epoch in the development of physics,
when the quanta and atoms were subjects of entertainment for the boys who
created the contemporary physics.

Similar experiments are given only in the best universities~\cite{Wash:12, MIT:13}
and whoever has an access to the set-ups' descriptions
may compare the accuracy with our high school experiment.

The theoretical subproblems are a good exercise in Ohm's law,
which give the opportunity for comprehensive understanding of the set-up's operation,
and are also a good starting point for understanding the electronics that
is the basis of many physical devices.
The experimental set-ups are gifts from the Olympiad organizers to the participants,
who can use them and make demonstrations in physics classes.
After the end of the school year, these set-ups should be given to the 
physics classroom in the school in which the participants study. 

\section{Initial easy tasks. S}
\begin{enumerate}
\item Measure the voltages of the four 9~V batteries and measure with maximal accuracy
the voltage of the 1.5~V battery.
\item Place the 1.5~V battery in its holder, connect it potentiometrically and measure the
interval of voltages which you get by rotating the axis of the potentiometer.
This will be the voltage source in the next tasks.
Reversing polarity changes the voltage sign.
\item Connect the 9~V batteries with the connection clips by buckling the electrodes..

\section{Analogue squaring. M}
\item \textbf{Attention! From this moment on there is a possibility to burn the 
integral circuits, if you connect the batteries improperly.}
Orient the set-up you work with so that both wires at the edge to be on the right side
and the sign ``COM'' to be on the right at the bottom.
Take a look at the schematics in Fig.~\ref{PCB}.
\item Carefully connect the 9~V female connector to the right hand side 3-pin male connectors on the circuit board, label to label.
Work carefully -- if you make an error with the polarity you will burn the integral circuits.
\item Connect the potentiometrically connected battery with the input wire of the set-up
at the top with a label ``IN'' and the other electrode of the potentiometer connect to
the wire comes from the ``ground'' of the circuit with a sign ``COM''.
\item Parallel to the potentiometer and ``IN''--``COM'' inputs of the circuit,
connect the first voltmeter, which shows voltage $U_1$.
\item At proper connectivity of the circuit, when you rotate the potentiometer axis,
the voltage $U_1$ should change approximately between 0 and the battery voltage 1.5~V.\item The second voltmeter, that shows voltage $U_2$, connect between the output wire of 
the circuit ``OUT'' and the common point ``COM''.
This way $U_2$ is the voltage between ``OUT'' and ``COM'' points.
\item Check whether the ``COM'' electrodes of both multimeters and the set-up are connected.
\item Rotate the potentiometer axis, wait 1 minute and write down the voltages $U_1$ and $U_2$.
Switch the polarity of the voltage source and repeat the measurements of the
input voltage $U_1$ and the output voltage $U_2$.
Arrange the results in a table with columns:
number of measurement $i$, $U_1$ and $U_2$. 
\item Represent the results in a graph in the plane $U_1$ abscissa (horizontal) and
$U_2$ ordinate (vertical).
\item Add a column to the table $(U_1)^2$ and represent the results graphically
in the plane $(U_1)^2$ abscissa and $U_2$ ordinate.
Draw a straight line that passes most closely to the experimental points.
This fitting (approximating) line is described by the equation
$U_2=(U_1)^2/U_0 + \mathrm{const}.$
Select two points on the straight line, measure the differences in abscissa 
$\Delta(U_1^2)$, and in ordinate $\Delta(U_2)$ 
and determine the parameter with dimension voltage from the slope
$U_0=\Delta(U_1^2)/\Delta(U_2)$.
This parameter of analogue multiplying schematics is essential for the determination of the
Boltzmann's constant $k_\mathrm{_B}$ and this is described in the subproblems of the next section.

\section{Voltage fluctoscopy. Determination of Boltzmann's constant. L} 
\begin{figure}
\includegraphics[scale=1.0]{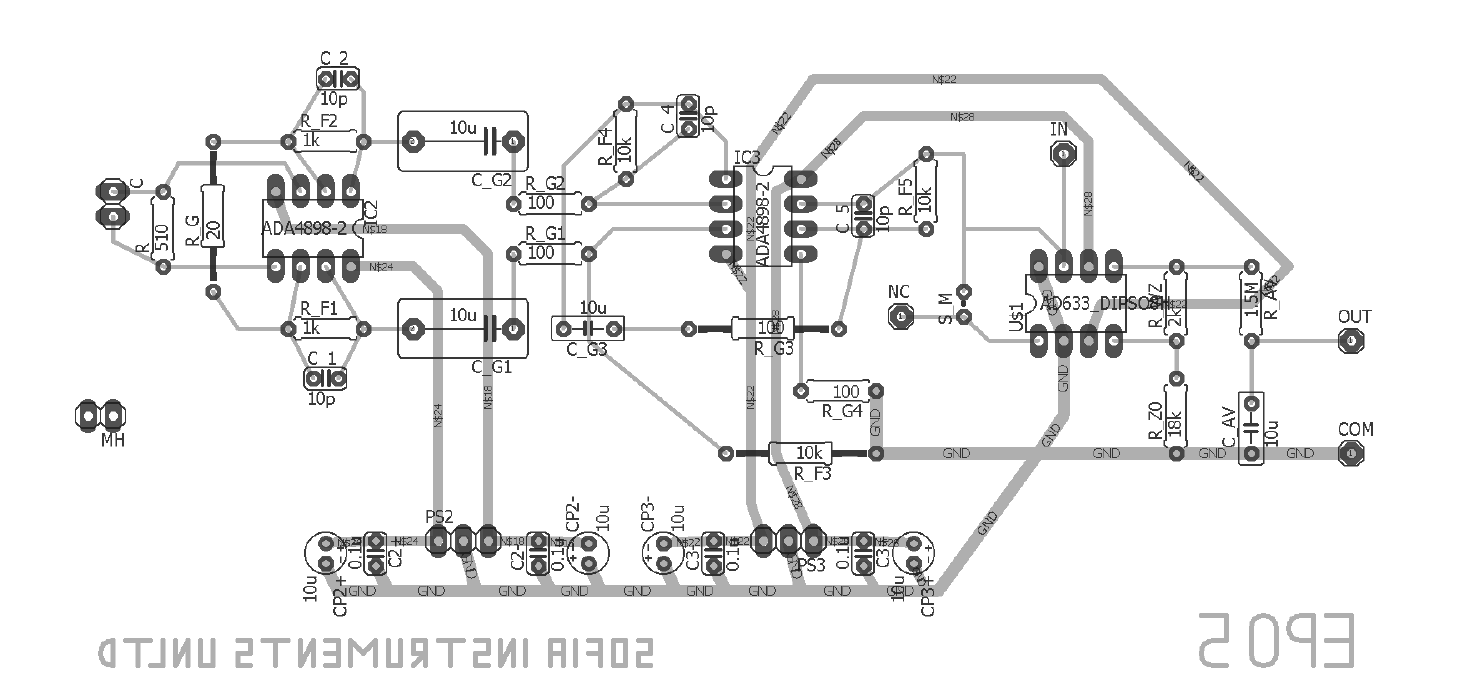}
\caption{PCB}
\label{PCB}
\end{figure}

\item Disconnect the potentiometer and the voltmeter, measuring the  ``IN''--``COM'' voltage $U_1$.  

\item Carefully connect the second voltage source with two 9~V batteries. Label to label, orange mark to orange mark, otherwise the integral circuit will be damaged.

\item This is the most dangerous moment for the integrated circuits (ADA4898-2). Mount both integrated circuits on the board. Again: green mark to green mark, label to label. If you make a mistake you will burn the integrated circuit.

\item You should have available a set of capacitors soldered to connectors.
The capacity $C$ of each capacitor is written in nanofarads ($\mathrm{1\;nF=10^{-9}\;F}$) on the label  glued to the connector. You can arrange the capacitors by the size of their capacity and write down those capacities $C_n$ in a table; the index $n$ shows the consecutive number.

\item What follows is the most important measurement -- determining the thermal fluctuations of the voltage of the capacitor. Keep the wire connected to the ``NC'' label of the set-up away from the power supplies. Attach in a succession each of the capacitors on the connector at the left end of the set-up. Wait for 2 minutes and write down the voltage $U_2$ shown by the voltmeter connected between the  ``OUT''  and ``COM'' electrodes of the set-up. Arrange the results in a table with columns $n$, $C_n$ and $U_2$.

\item The fluctuating voltage $U$ of the capacitor $C$ is amplified one million times, more precisely, the gain coefficient is $Y=1.01\times 10^6$, compare that to the amplifier of the Habicht brothers.~\cite{Habicht:10}
Calculate the parameter $U^*=U_0/Y^2$. It has dimension of voltage and is important for the experiment. Do not be alarmed by the powers, this is a really small voltage of the order of picovolt ($\mathrm{1\;pV=10^{-12}\;V}$).

\item Evaluate the room temperature and write it down in Kelvins (0~K$\; \approx -273^\circ$C).

\item Add two new columns to the table:
$y_n= U_{2,n} U^* /T$ and $x_n=1/C_n$;
This is the most important experimental data!

\item The results in the table should be represented visually by a graph in the $x$-$y$ plane. Think for a few moments what the proper scaling should be. 

\item If you have worked correctly the individual points should be close to a straight line. Draw the line $y=k_\mathrm{_B}\,x + \mathrm{const}$, which you think is closest to all the points.
This is a standard mathematical procedure, called linear regression and is implemented numerically in a lot of calculators so you may, if you want, get the solution numerically.

\item Choose two points from the line and determine the difference in their coordinates on the $y$ axis - $\Delta y$, and the $x$ axis - $\Delta x$.
Write down those differences.

\item Calculate the slope of the line $k_\mathrm{_B}=\Delta y/\Delta x$ and draw a box around the result.
Congratulations, you have just measured a fundamental constant - Boltzmann's constant $k_\mathrm{_B}$! Just determining the order correctly is a big success.

\section{Theoretical problems for Ohm's law related to the experimental set-up. XL}

This is an Olympiad in experimental physics, solving the problems in this section will give you only a few points, and their impact on the final score may be insignificant.

The results from the solution of those problems will be used only for differentiating works in which the experiments were performed equally well, or for non-zero results in cases in which the participant has burnt the integrated circuits by not following correctly the assembly instructions.

The circuits shown in Figures~
\ref{Non-inverting amplifier}, \ref{Buffer}, \ref{Differential amplifier} and \ref{Inverting amplifier}
contain triangles. The left vertical side of each triangle has inputs, denoted by (+) and (-). The output is on the right and is denoted by (0). The output current $I_0$ is such that the input voltages equalize $U_+=U_-$. The currents at the inputs (+) and (-) are negligible. Such device is called an operational amplifier.

\item By using the Ohm's law show that the relation between voltages and currents is given by the equation $y_1=V_2/V_1=1+R_f/r_g$,  Fig.~\ref{Non-inverting amplifier}.

\item Analogously, for the symmetric variant of the same circuit
$y_1=(V_2-V_4)/(V_1-V_3)=R_f/r_g+1$, Fig.~\ref{Buffer}.

\item $y_2=V_7/(V_5-V_6)=-R_f/r_g$. Fig.~\ref{Differential amplifier}.

\item $y_3=V_9/V_8 =-R_f/r_g$. Fig.~\ref{Inverting amplifier}.

\item If a large capacitor $C_g$ is connected in series with a resistor $r_g$ show that, for frequencies such that $\omega r_g C_g \gg 1$, the effect of the capacitor is negligible.

\item For which frequencies, we can ignore the effect of a small capacitor $C_f$, connected in parallel with a large resistor $R_f$?

\item If the three amplifiers, are connected in series, the total gain is  $Y = y_1 y_2 y_3$. Make an estimate for the frequency interval $(f_\mathrm{l},\,f_\mathrm{h})$, for which the gain coefficient $Y$ is approximately constant. Use the numerical values of the parameters in table~\ref{tbl:values}; $f=\omega/2\pi$

\item In the circuits in Fig.~\ref{Analog multiplier} the voltage $U_W=U_X U_Y/U_m+U_Z$, with $U_X=U_Y$, and the constant $U_m=10\;\mathrm{V}$ for the used amplifier AD633. Express $U_Z$ through $U_W$ and show that $U_W=U_X^2/\tilde{U}$, where $\tilde{U}=U_\mathrm{m}\,R_1/(R_1+R_2)$.

\item Show that  $U_2=R_\mathrm{V} \left<U_W\right>/(R_\mathrm{av}+R_\mathrm{V})$, where $R_\mathrm{V}$, most frequently $1\;\mathrm{M}\Omega$, is the internal resistance of the ohmmeter, and the brackets $\left<U_W\right>$ denote time average.

\item Show that $U_2=U_X^2/U_0$, where $U_0=\tilde U (R_\mathrm{av}+R_\mathrm{V})/R_\mathrm{V}$.

\item 
Finally, the main equation for our set-up is 
$U_2=\left< U^2\right>/U^{*}$,
where $U$ is the capacitor voltage at the input of the device,
and
\be
\frac1{U^*} \equiv \frac{Y^2}{U_0}
=\frac{Y^2}{U_m}\frac{R_1+R_2}{R_1}\frac{R_\mathrm{V}}{R_\mathrm{av}+R_\mathrm{V}}.
\ee

\item Calculate the values of the parameter $U^*$ by using the parameters in Table~\ref{tbl:values}.

\begin{figure*}[h]
\includegraphics[scale=0.25]{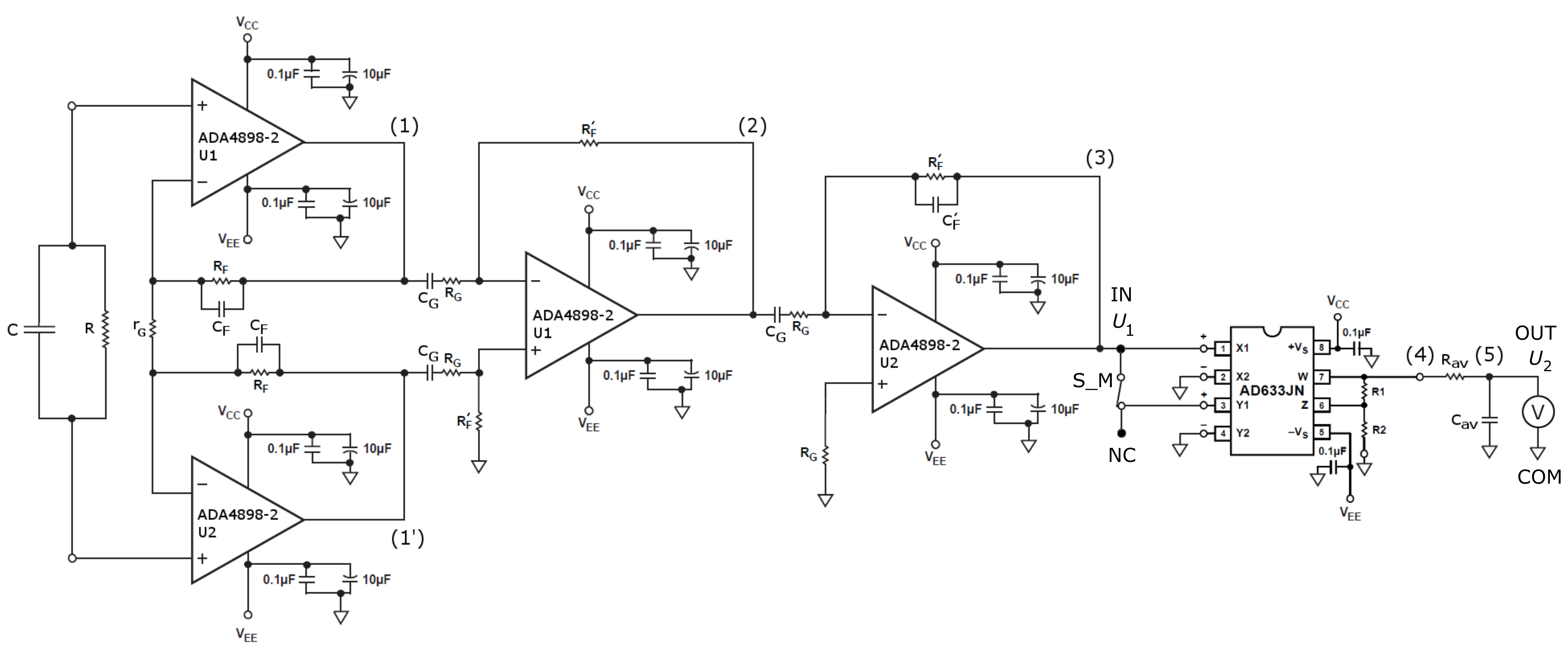}
\caption{Circuit.}
\label{fig:circuit}
\end{figure*}

\begin{center}
\begin{table}[h]
\begin{tabular}{| c | r |}
		\hline
		&  \\ [-1em]
		Circuit element  & Value  \\ \tableline
			&  \\ [-1em]
			$R$ & 510~$\Omega$ \\
			$r_\mathrm{_G}$ & 20~$\Omega$ \\
			$R_\mathrm{F}$ &  1~k$\Omega$  \\
			$C_\mathrm{F}$ &  10~pF  \\ 
			$C_\mathrm{G}$ & 10~$\mu$F \\
			$R_\mathrm{G}$ &  100~$\Omega$  \\ 
			$R_\mathrm{F}^\prime$ & 10~k$\Omega$ \\
			$C_\mathrm{F}^\prime$ & 10~pF \\
			$R_1$ &  2~k$\Omega$  \\ 
			$R_2$ & 18~k$\Omega$  \\
			$R_\mathrm{av}$ & 1.5~M$\Omega$ \\
			$C_\mathrm{av}$ & 10~$\mu$F \\
			$R_\mathrm{_V}$ & $\approx 1~\mathrm{M} \Omega$ \\
			$V_\mathrm{CC}$ & +9~V \\
			$V_\mathrm{EE}$ & -9~V \\
\tableline
\end{tabular}
	\caption{Table of the numerical values of the circuit elements from Fig.~\ref{fig:circuit}.}
	\label{tbl:values}
\end{table}
\end{center}

\end{enumerate}

\begin{figure}[t]
\begin{minipage}[t]{0.31\linewidth}
\includegraphics[scale=0.28]{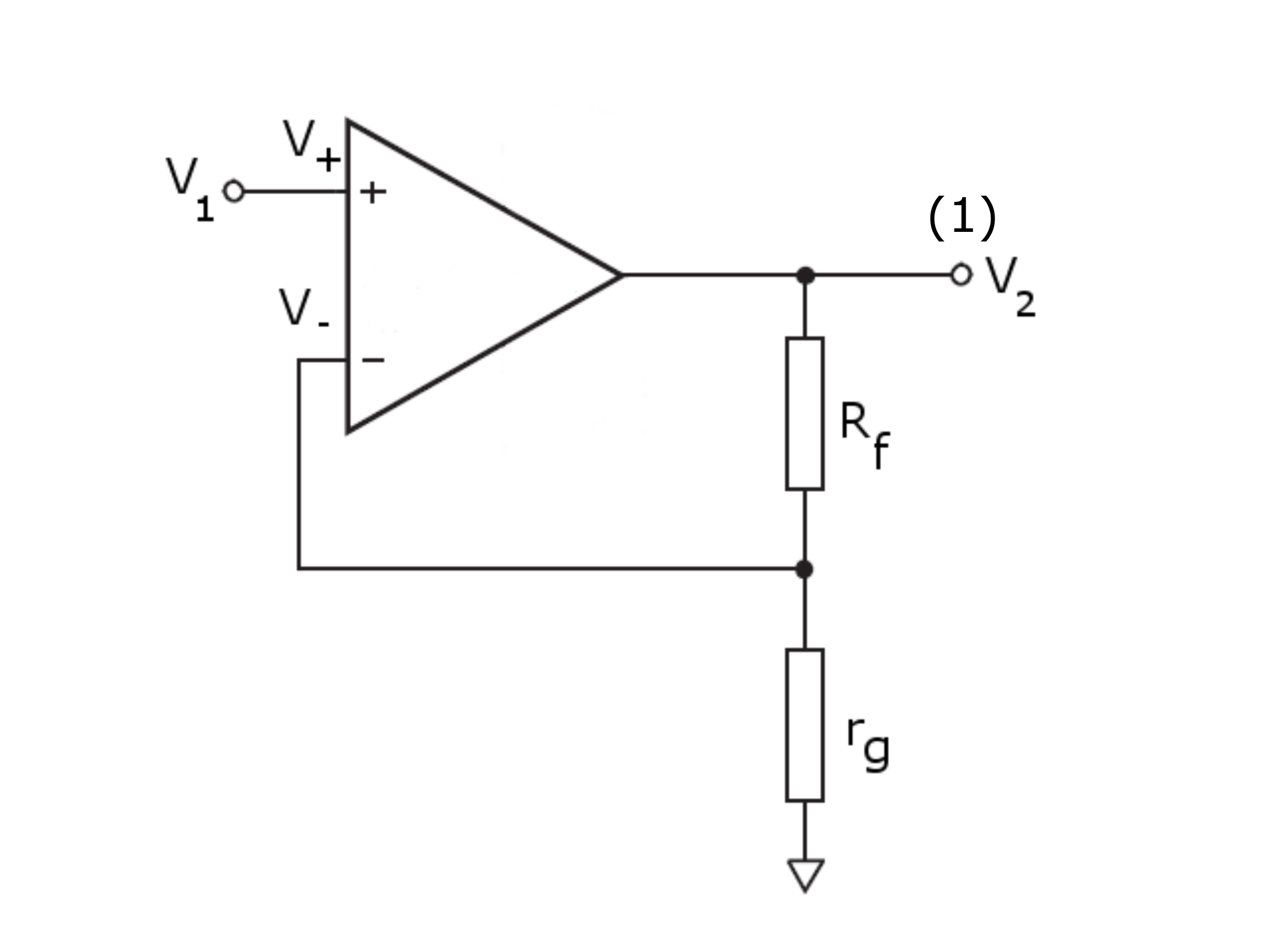}
\caption{Non-inverting amplifier}
\label{Non-inverting amplifier}
\end{minipage}
\begin{minipage}[t]{0.31\linewidth}
\includegraphics[scale=0.28]{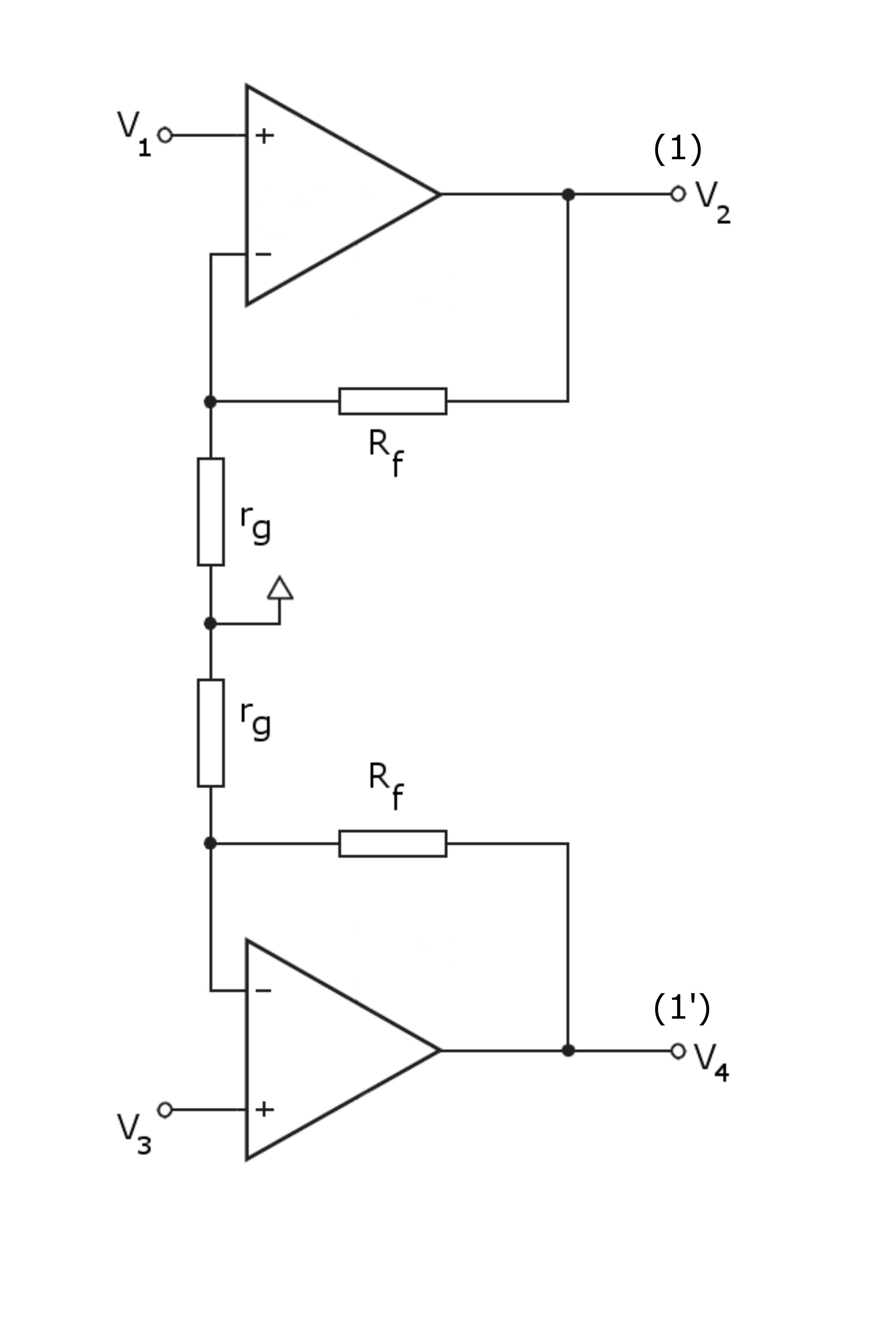}
\caption{Buffer}
\label{Buffer}
\end{minipage}
\begin{minipage}[t]{0.36\linewidth}
\includegraphics[scale=0.28]{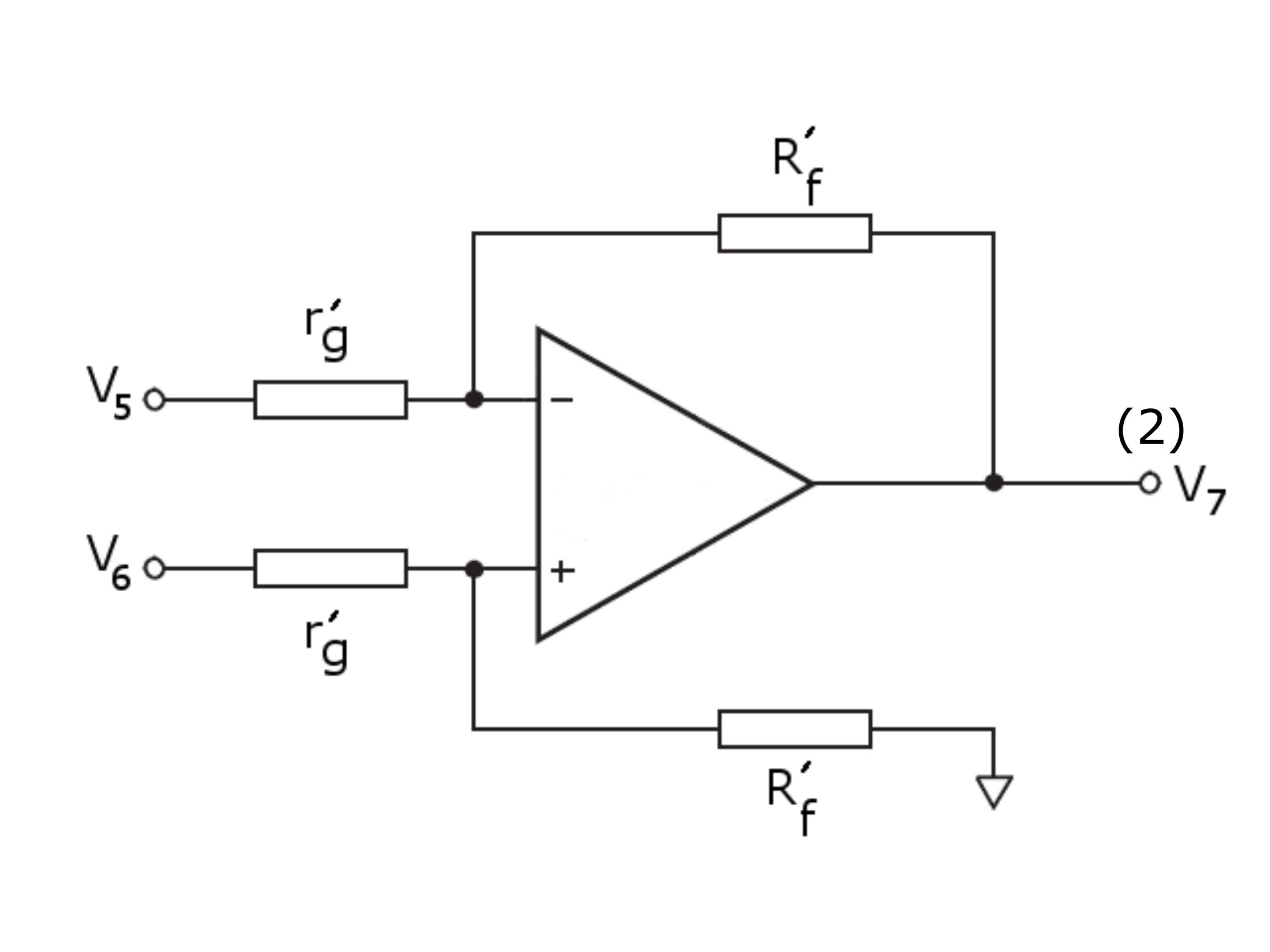}
\caption{Differential amplifier}
\label{Differential amplifier}
\end{minipage}
\begin{minipage}[c]{0.4\linewidth}
\includegraphics[scale=0.28]{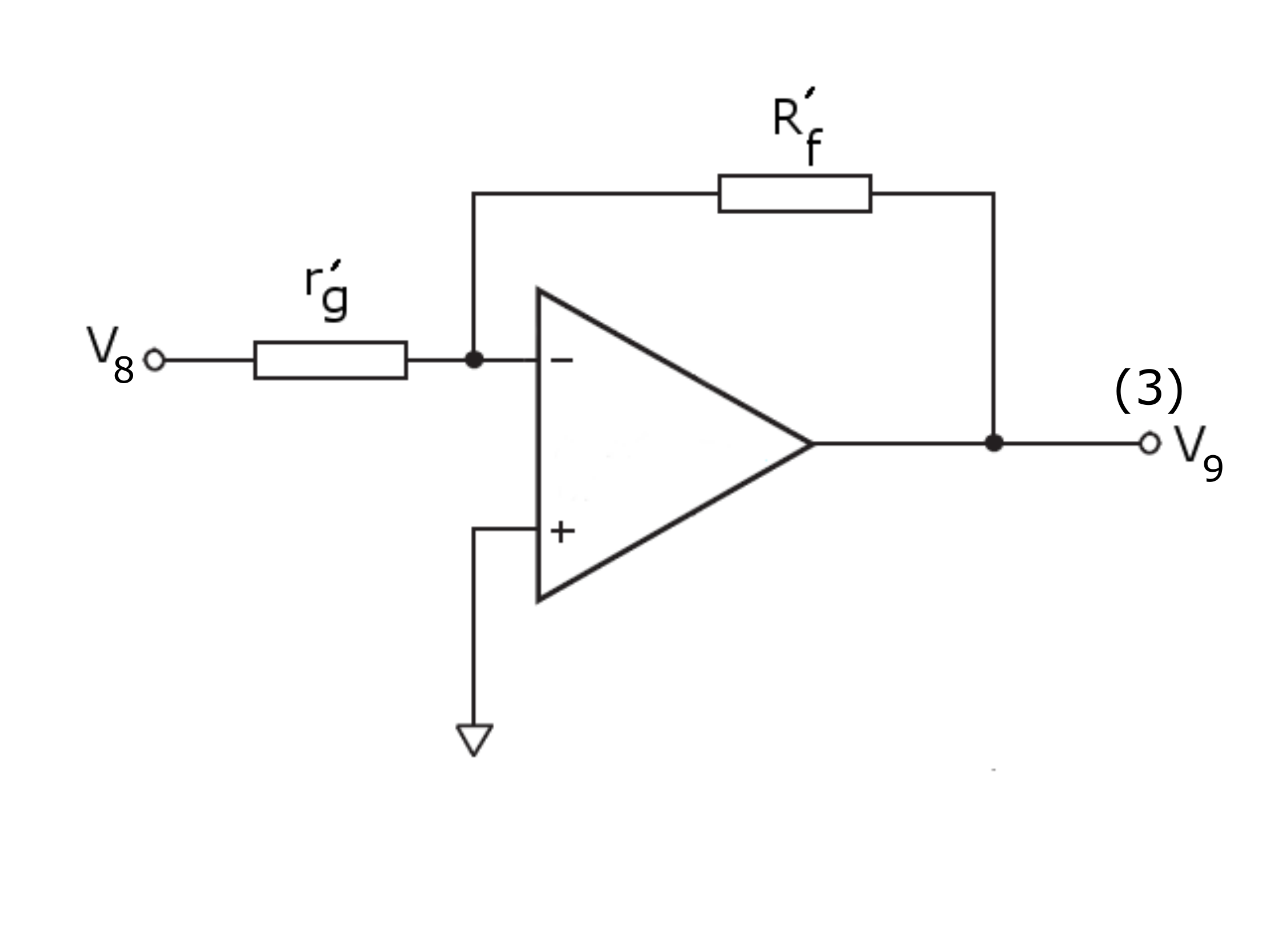}
\caption{Inverting amplifier}
\label{Inverting amplifier}
\end{minipage}
\begin{minipage}[c]{0.57\linewidth}
\includegraphics[scale=0.28]{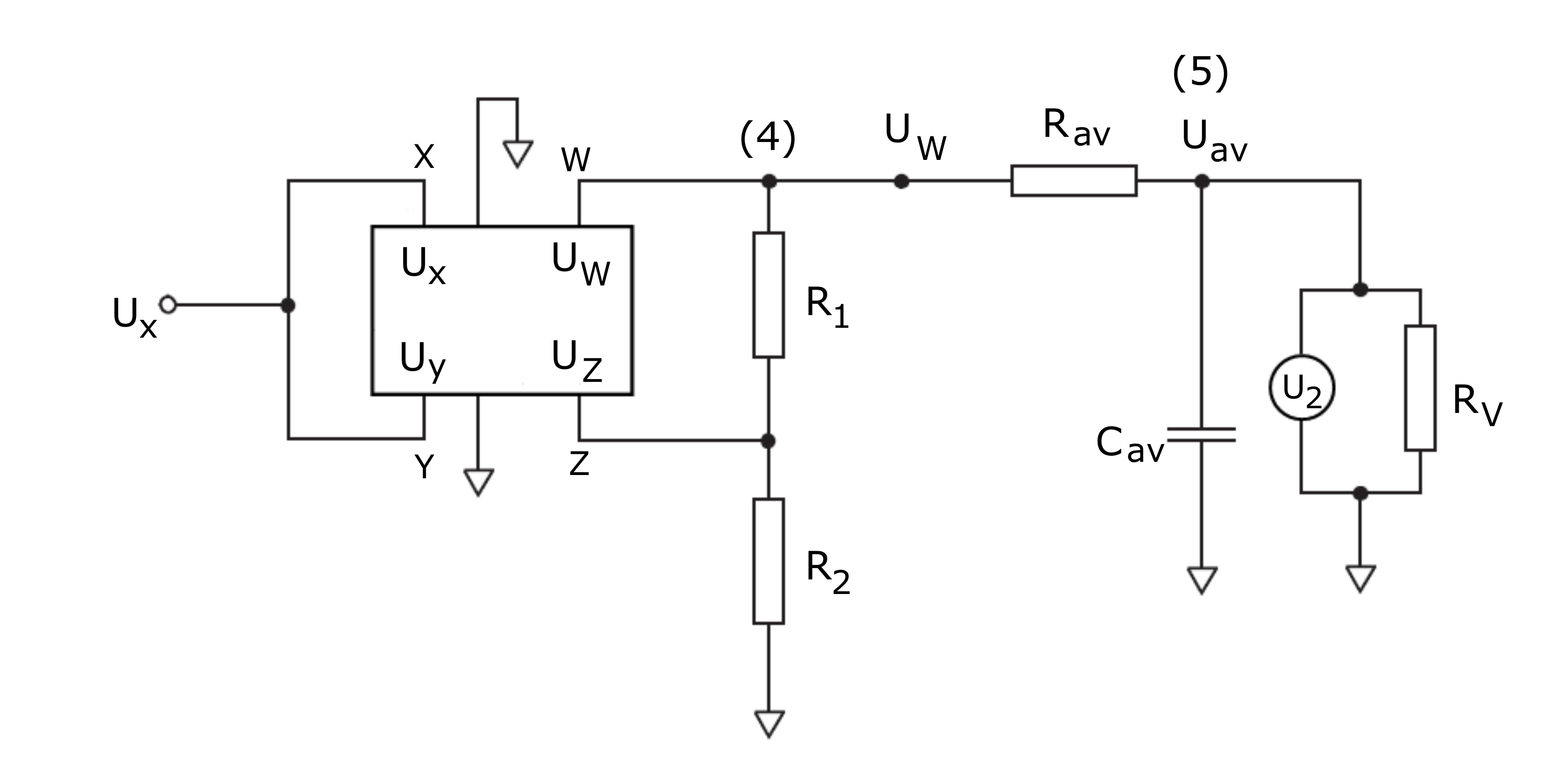}
\caption{Analog multiplier}
\label{Analog multiplier}
\end{minipage}
\end{figure}

\section{Homework problem, the solution to which must be send during the night after the Olympiad, by sunrise, at the e-mail of the Olympiad.
Derive the formulas used for determining Boltzmann's constant $k_\mathrm{_B}$.
Sommerfeld reward with a monetary equivalent of DM137. XL}

Instructions. Start with the equipartition theorem applied to the energy of the temperature fluctuations.
$\frac12 C\left< U^2 \right>=\frac12 k_\mathrm{_B}T.$
By analyzing the electronic circuit of the set-up, express the mean-square of voltage 
$\left<U^2\right>=U_2 U^* $ by the experimentally measured voltage $U_2$.
Express the coefficient $U^*$ in the linear regression formula 
$U_2 U^* /T= \left<U^2\right>/T = k_\mathrm{_B}\, (1/C) + \mathrm{const},$
from which, by the slope of the line, you can determine the Boltzmann's constant $k_\mathrm{_B}$.
In a few words, write an article showing how the idea of Einstein for determining Boltzmann's constant is realized with our set-up.
Show what improvements to the set-up would increase the accuracy of the experiment.

You can work as a group with other participants in the Olympiad, you can also consult with professionals, but you must give the names of the consultants and the full list of authors.

The best article will be rewarded the a Sommerfeld reward with a monetary equivalent of DM137.
The prize is given personally and only in the day in which the results are announced - 10.12.2017.

\section{Problems for further work}

The Olympiad is created with the idea of helping students, deprived of quality education, students coming from schools where the physics classrooms are closed.
For this reason, do not get upset if you are not satisfied with the achieved results.
Ask your teacher if there is an appropriate measuring apparatus for measuring the capacitors more precisely. Do the same experiment calmly in your home and you will see that it is simple.
If you can measure a fundamental constant in your home, then you have a flying start for all natural and technical sciences.
If you have the opportunity, say in your physics club, demonstrate the set-up in your school.

Please, write us a short text containing your opinion about the Olympiad.
We highly value your recommendations, and will try to take them into account for the next issue of the Olympiad - EPO6, next year at the same time.

The experimental set-up is indeed unique.
You have in your hands an amplifier, capable of amplifying a million times, with a noise level of $1\;\mathrm{nV/\sqrt{Hz}}.$

With the proper modification the set-up can be used for measurement of the electron charge and the absolute temperature. Furthermore, with a minimal modification, the device can be used as a lock-in voltmeter, which can measure alternating voltages even smaller then 1 microvolt ($\mathrm{1\;\mu V=10^{-6}\;V}$). For those reasons, there is an output, denoted by ``NC'', which is not used in the problems for the Olympiad.
For more details, please follow the publications of the authors, on the server, containing the problems and their solutions from the previous issues of the Olympiad.

\appendix

\section{A photograph of the experimental set-up}

\begin{figure}[h]
\centering
\includegraphics[scale=0.25]{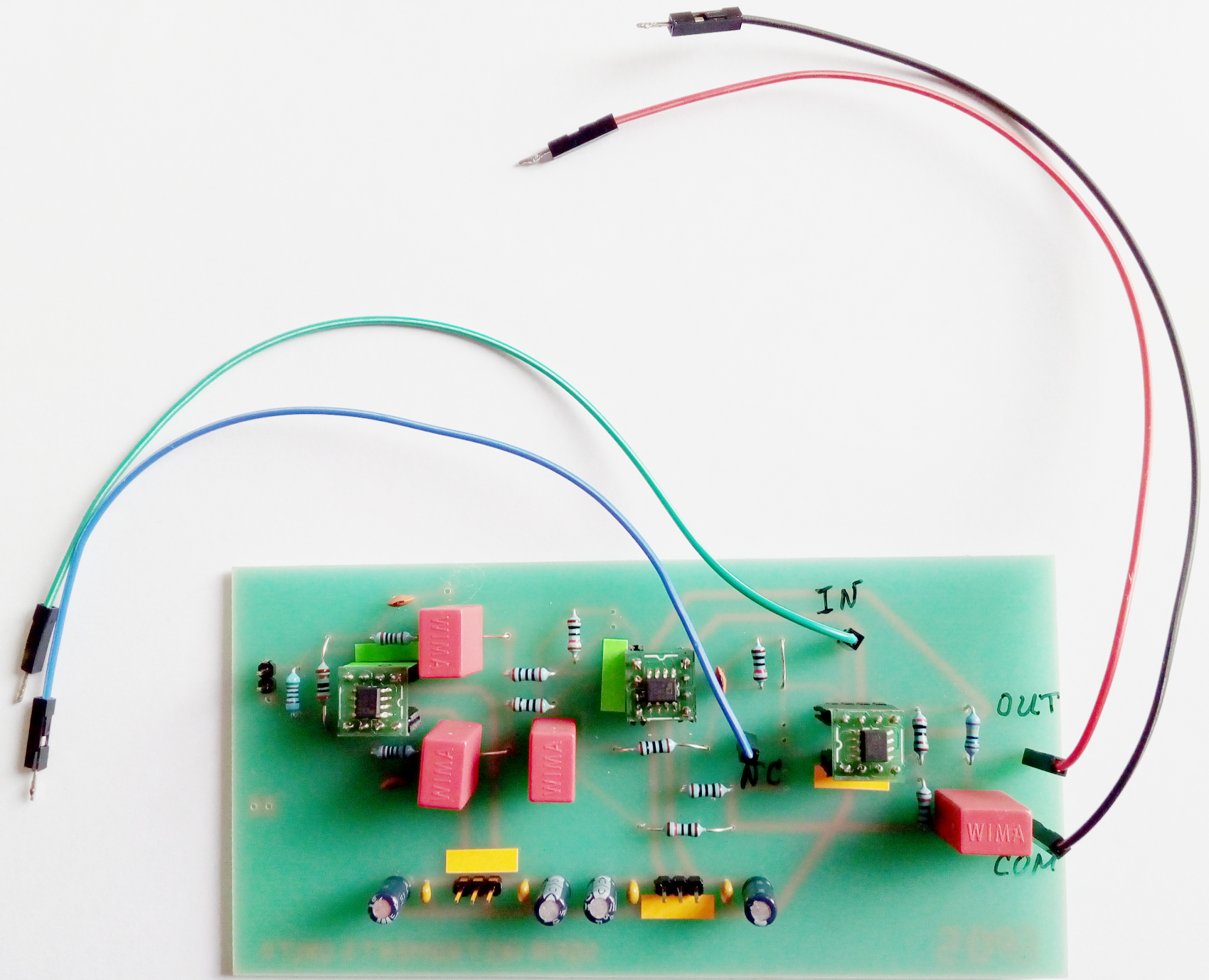}
\caption{A photograph of the experimental set-up.
The electrodes and wires between which the difference $U_2$ of the output voltage ``OUT'' and the ground ``COM'' are visible on the right.
The input voltage $U_1$ is applied between the input ``IN'' and the ground ``COM''.
The wire ``NC'' can be used for further research.
On the 2-pin male connector on the left at the end the different capacitors $C_n$ are placed.}
\label{Fig:Photo}
\end{figure}

\section{Solution to the experimental problems}

\subsection{Initial easy tasks. S}
\begin{enumerate}
\item $U_{\mathrm{B}1}=9.72$~V, $U_{\mathrm{B}2}=9.73$~V, $U_{\mathrm{B}3}=9.73$~V, $U_{\mathrm{B}4}=9.73$~V and for the 1.5~V battery measurement the multimeter range that gives the best accuracy should be used (usually this range is 2~V),
$U_{\mathrm{B}5}=1.582$~V.
\item The potentiometer has three terminals, the outer two of which are soldered to the battery holder.
A potentiometer connectivity means connection to the middle and one of the outer terminals,
$U_\mathrm{min}=-1.577$~V, $U_\mathrm{max}=1.577$~V and therefore the voltage interval is 
$U \epsilon [-1.577,1.577]$~V.
\item Simply connecting the 9~V batteries to their connection clips.

\subsection{Analogue squaring. M}

\item Orient the experimental set-up according to the instructions.

\item Turn on the voltage according to the instructions.

\item Connect the battery to the described in the instructions electrodes of the experimental set-up.

\item Connect a voltmeter parallel to the input according to the instructions.

\item A check for proper connectivity.

\item Connect the second voltmeter to the output of the set-up according to the instructions.

\item A check for proper connection of both voltmeters, the ``COM'' electrodes of which should be connected.

\item Table~\ref{tbl:mult} without the last column (task 13).
\begin{center}
\begin{table}[h]
\begin{tabular}{ c  r  r  r }
		\tableline \tableline
		&  \\ [-1em]
		i & $U_1$ [V] & \hspace{2.5pt} $U_2$ [V] & \hspace{5pt} $U_1^2$ [V$^2$] \\ \tableline 
			&  \\ [-1em] 
			1	&	0.1	&	0.011	&	0.01	\\
			2	&	0.2	&	0.023	&	0.04	\\
			3	&	0.4	&	0.07	&	0.16	\\
			4	&	0.6	&	0.15	&	0.36	\\
			5	&	0.8	&	0.263	&	0.64	\\
			6	&	1.0	&	0.386	&	1.00	\\
			7	&	1.2	&	0.551	&	1.44	\\
			8	&	1.4	&	0.748	&	1.96	\\
			9	&	1.573  &	0.941	&	2.474	\\
			10	&	-1.572	&	0.942	&	2.471	\\
			11	&	-1.4	&	0.75	&	1.96	\\
			12	&	-1.2	&	0.553	&	1.44	\\
			13	&	-1.0	&	0.387	&	1.00	\\
			14	&	-0.8	&	0.265	&	0.64	\\
			15	&	-0.6	&	0.151	&	0.36	\\			
			16	&	-0.4	&	0.071	&   0.16	\\
			17	&	-0.2	&	0.024	&	0.04	\\
			18	&	-0.1	&	0.012	&	0.01	\\			
\tableline \tableline
\end{tabular}
	\caption{Results from the measurements in task 11 and the data processing in task 13 (last column).}
	\label{tbl:mult}
\end{table}
\end{center}
\item The results are presented in Fig.~\ref{fig:par}.
\begin{figure}[h]
\begin{minipage}[t]{0.48\linewidth}
\centering
\includegraphics[scale=0.64]{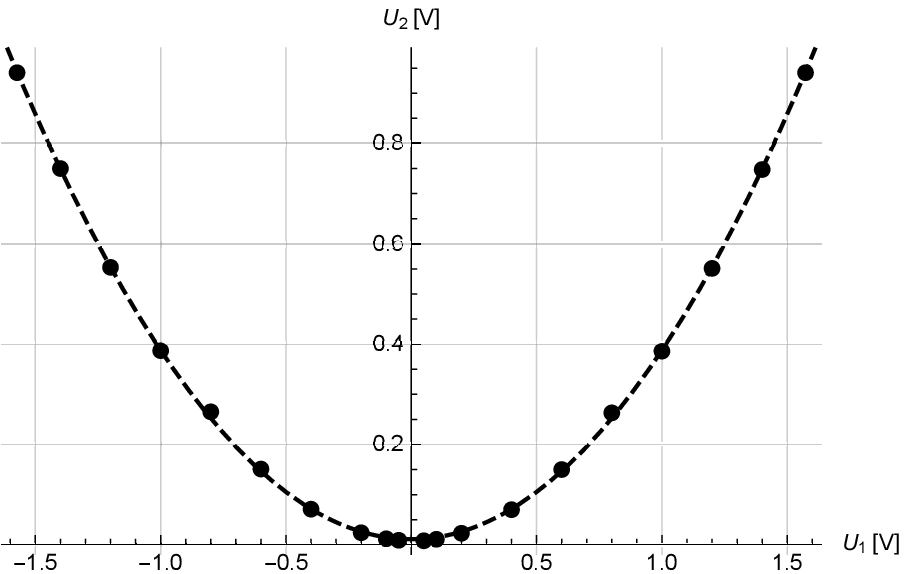}
\caption{The graphically represented dependency $U_2$ of $U_1$ from Table~\ref{tbl:mult}.}
\label{fig:par}
\end{minipage}
\hfill
\begin{minipage}[t]{0.48\linewidth}
\centering
\includegraphics[scale=0.64]{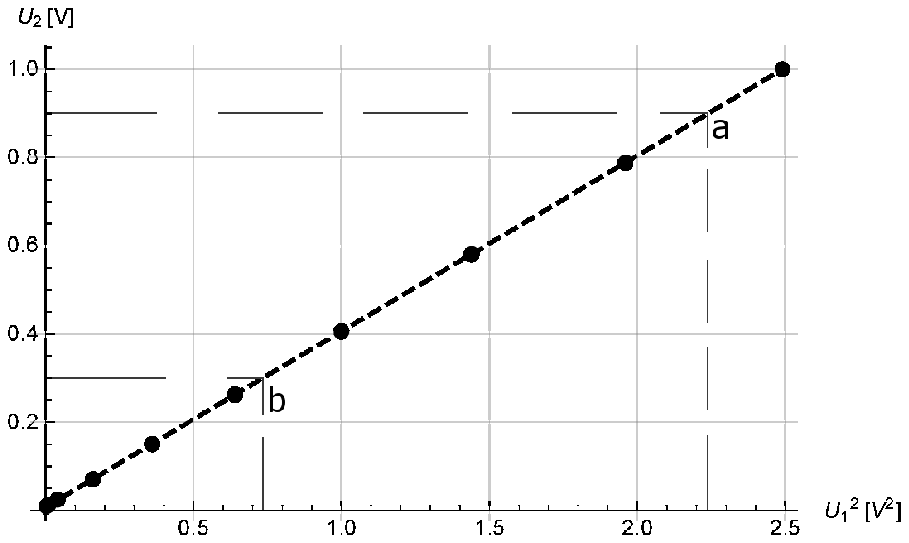}
\caption{The graphically represented dependency $U_2$ of $(U_1)^2$ from Table~\ref{tbl:mult}.}
\label{fig:lin}
\end{minipage}
\end{figure}
\item The last column of Table~\ref{tbl:mult} and the results presented in Fig.~\ref{fig:lin}.
We choose points (a) and (b) in Fig.~\ref{fig:lin}, so that in ordinate  $(U_2)_a=0.9$~V and , $(U_2)_b=0.3$~V and we find their corresponding values in abscissa $(U_1)_a^2 \approx 2.24$~V$^2$  and $(U_1)_b^2 \approx 0.725$~V$^2$ by for each point drawing a line parallel to the ordinate towards the abscissa, and the value of the intersection point is the corresponding $(U_1)^2$
\be
U_0=\frac{\Delta(U_1^2)}{\Delta(U_2)}=
\frac{(U_1)_a^2-(U_1)_b^2}{(U_2)_a-(U_2)_b}=\frac{2.24-0.725}{0.9-0.3}=2.525~\mathrm{V}.
\ee

\subsection{Voltage fluctoscopy. Determination of Boltzmann's constant. L} 

\item Disconnect the input voltage source and input voltmeter according to the instructions.

\item Apply voltage to the amplifier \textbf{according to the instructions}.

\item Connect the two operational amplifiers to the printed circuit board \textbf{according to the instructions}.

\item The capacity values of the capacitors set are written in the first 2 columns of Table~\ref{tbl:meas}.
\begin{center}
\begin{table}[h]
\begin{tabular}{ c  r  r  r  r }
		\tableline \tableline
		&  \\ [-1em]
		$n$  \hspace{1.5pt} & $C_n$ [nF] & \hspace{5pt} $U_2$  [V] & \hspace{1.5pt} $x_n$ [1/$\mu$F] & \hspace{1.5pt} $y_n$ [f\,V$^2$/K]\\ \tableline 
			&  \\ [-1em] 
			1	&	14.8	&	0.293	&	67.57 	&	2.458 \\
			2	&	27.9	&	0.255	&	35.84	&	2.139 \\
			3	&	33.4	&	0.244	&	29.94	&	2.047 \\
			4	&	42.3	&	0.233	&	23.64	&	1.955 \\
			5	&	66.9	&	0.202	&	14.95	&	1.695 \\
			6	&	105.0	&	0.204	&	9.524	&	1.712 	\\
\tableline \tableline
\end{tabular}	\caption{Experimental results from the measurement of the voltage of the thermal fluctuations for the set of capacitors. Here for brevity we have included the prefix $\text{f}=10^{-15}$.}
	\label{tbl:meas}
\end{table}
\end{center}
\item The results from the measurement according to the instructions are written in the 3-rd column of Table~\ref{tbl:meas}.

\item For $U_0$ we obtained 2.525~V, therefore
\be
U^*=\frac{U_0}{Y^2}=\frac{2.525}{(1.01 \times 10^6)^2}=\frac{2.525}{(1.01)^2 \times 10^{12}}=
2.475 \times 10^{-12}=2.475~\mathrm{pV}.
\ee

\item No thermometer is necessary for the evaluation (not exact measurement) of the temperature in the auditorium.
For room temperature values between 20$^\circ$C and 25$^\circ$C are acceptable, in Kelvins (we add  273$^\circ$C more) respectively 293~K and 298~K, we will pick up the lesser value $T=293$~K.
The difference between both values is below 2\%, which is completely satisfying for our problem.
Moreover, these 5$^\circ$C are fully sensible and in practice any human being is capable to feel even smaller temperature differences (for instance whether it is warm enough in the auditorium for one to work in a t-shirt).
Your body is a thermometer and like any measuring device, it needs proper calibration for reliable readings. 

\item The results are written in the last 2 columns of Table~\ref{tbl:meas}, where the second index $n$ of the notion $U_{2,n}$ is the number of the respective capacitor and the prefix ``f'' for fV$=10^{-3}~\mathrm{pV}=10^{-15}$~V is called femto-volt.

\item Assess the scale for the graphical representation of the obtained measurements.

\item The straight line is drawn in Fig.~\ref{fig:uc}.

\begin{figure}[h]
\centering
\includegraphics[scale=1.0]{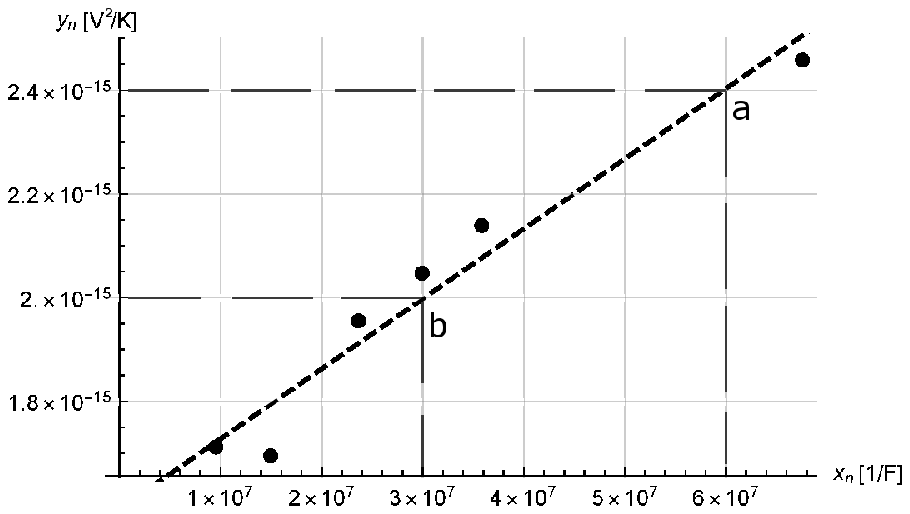}
\caption{Graphical representation of the dependency $y_n$ of $x_n$ from Table~\ref{tbl:meas}.
First, the experimental points are placed.
Then an approximating line is drawn.
This line can be drawn with an eye, while the mathematical procedure is called a linear regression.
On the straight line we choose 2 points (a) and (b) and calculate the slope
$k_\mathrm{_B}=\Delta y/ \Delta x$, which in our case gives the Boltzmann constant.}
\label{fig:uc}
\end{figure}

\item Analogously to problem 13, we choose 2 points on the straight line (a) and (b) with respective coordinates $x_a= 6 \times 10^7~$F$^{-1}$, $y_a=2.4 \times 10^{-15}$~V$^2$/K and $x_b= 3 \times 10^7~$F$^{-1}$, $y_b=2 \times 10^{-15}$~V$^2$/K and we calculate the difference in their coordinates 
\begin{align}
& \Delta y=y_a-y_b=(2.4-2) \times 10^{-15} = 0.4 \times 10^{-15} \, \mathrm{V^2/K}, \\ 
& \Delta x=x_a-x_b=(6-3) \times 10^7 = 3  \times 10^7 \, \mathrm{1/F}.
\end{align}

\item With the obtained values for the coordinate differences, we calculate the slope of the line, which in fact is $k_\mathrm{_B}$
\be
k_\mathrm{_B}=\frac{\Delta y}{\Delta x}=
\frac{0.4}{3} \times 10^{-22}=1.33 \times 10^{-23} \,( \mathrm{F\,V^2/K}=\mathrm{J/K}).
\ee

Let us account that the Boltzmann constant is very small and even the correct determination of its magnitude with such a simple set-up is a significant success.
When the signal is amplified a million times in voltage and
$10^{12}$ times (120~dB) in power
a multiplier of 2 (3~dB) means
an error in logarithmic scale 3/120=2.5~\%,
which is a considerable success for an experimental set-up for the high school education.
Although, it happens the set-up to give an accuracy of several percent.
With the well-known experimental value
$k_\mathrm{_B}=1.38064 \times 10^{-23}$~J/K
the relative measurement error is
$(138-133)/138<4\%.$
The effects of non-ideality of the operational amplifiers give systematic errors,
but the main source of uncertainty is the floating of the zero of the experimental set-up.
The reduction of the floating of the zero is possible but this would significantly complicate the circuit up to the level of university physics or even metrology.

\end{enumerate}

\section{Several simple tasks on Ohm's law}

Let us consider 6 consecutive tasks that are solved by Ohm's law application.
In Figures~\ref{Non-inverting amplifier}-\ref{Inverting amplifier}  the current going out from the right angle of the triangle is such that the voltages at the points  (+) and (-) are equal
\begin{equation}
V_+-V_-=0,
\label{eq:poten}
\end{equation}
and the currents in these inputs are negligible.
\begin{enumerate}
\item Let us study the current $I_2=V_2/(R_\mathrm{f}+r_\mathrm{g})$, which goes down from the output  $V_2$ to the ``ground'' 
(\begin{circuitikz}
\draw (0,0) node[sground,xscale=0.25,yscale=0.45]{};
\end{circuitikz})
with potential 0.
According to Ohm's law the voltage at the point between the resistors $R_\mathrm{f}$ and $r_\mathrm{g}$, which is the same as it is at the (-) input is
\begin{equation}
V_-=r_\mathrm{g} I_2 = \frac{r_\mathrm{g}}{R_\mathrm{f}+r_\mathrm{g}}V_2.
\end{equation}
On the other hand, for each of the triangles the two voltages are nearly equal and
\begin{equation}
V_1=V_+=V_-=\frac{r_\mathrm{g}}{R_\mathrm{f}+r_\mathrm{g}}V_2.
\end{equation}
In this way we obtain for the gain coefficient
\begin{equation}
y_1=\frac{V_2}{V_1}=\frac{R_\mathrm{f}}{r_\mathrm{g}}+1.
\end{equation}
Because $y_1>0$ this amplifier is called non-inverting.
The electronic devices that equalise the potentials at both input terminals according to Eq.~(\ref{eq:poten}) are called operational amplifiers but the terminology only repels the beginners.
\item In Fig.~\ref{Buffer} two such amplifiers are drawn, for which 
\begin{align}
& V_2=y_1 V_1, \nn \\
& V_4=y_1 V_2.
\end{align}
If we subtract both equations, we obtain the same gain coefficient
\begin{equation}
\frac{V_2-V_4}{V_1-V_2}=y_1=\frac{R_\mathrm{f}}{r_\mathrm{g}}+1.
\end{equation}
The ground in Fig.~\ref{Buffer} is wrongly drawn because it is not necessary this point to have zero potential and it is left disconnected.
Such a ground is called virtual ground and instead of two serial connected resistors $r_\mathrm{g}$, one double resistor $r_\mathrm{_G}=2 r_\mathrm{g}$ is used and then
\begin{equation}
y_1=1+\frac{2 R_\mathrm{f}}{r_\mathrm{_G}}.
\end{equation}
Such a circuit with two non-inverting amplifiers is called a buffer.
The circuit parameters of the buffer drawn in Fig.~\ref{fig:circuit} are given in Table~\ref{tbl:values}, where
$r_\mathrm{_G} =20~\Omega, R_\mathrm{f}\equiv R_\mathrm{F}=1~\mathrm{k}\Omega$ and $y_1 \approx 101$.
We use 1\% accuracy resistors.
\item The amplifier shown in Fig.~\ref{Differential amplifier} is analysed by studying of voltage dividers described by Ohm's law, too.

From a point with potential $V_6$ to the ground a current goes down
$I_6=V_6/(R^\prime_\mathrm{f}+r^\prime_\mathrm{g})$.
According to Ohm's law the potential at point (+) is
\be
V_+=0+R^\prime_\mathrm{f} I_6 = \frac{R^\prime_\mathrm{f}}{R^\prime_\mathrm{f}+r^\prime_\mathrm{g}} V_6.
\label{eq:d-}
\ee
Completely analogically from point $V_5$ to $V_7$ a current $I_5=(V_5-V_7)/(R^\prime_\mathrm{f}+r^\prime_\mathrm{g})$ goes down and the potential at point (-) is
\be
V_-=V_7+\frac{V_5-V_7}{R^\prime_\mathrm{f}+r^\prime_\mathrm{g}}R^\prime_\mathrm{f}.
\label{eq:d+}
\ee
The current from the triangle (symbolising the operational amplifier) is such that $V_+=V_-$ and the substitution of the both upper Eqs.~(\ref{eq:d-} and \ref{eq:d+}) gives
\be
\frac{R^\prime_\mathrm{f}}{R^\prime_\mathrm{f}+r^\prime_\mathrm{g}} V_6=
V_7\frac{R^\prime_\mathrm{f}+r^\prime_\mathrm{g}}{R^\prime_\mathrm{f}+r^\prime_\mathrm{g}} +\frac{V_5-V_7}{R^\prime_\mathrm{f}+r^\prime_\mathrm{g}}R^\prime_\mathrm{f}.
\label{eq:d=}
\ee
From the right hand side of Eq.~(\ref{eq:d=}) the nominator terms that have $R^\prime_\mathrm{f} V_7$ cancel each other.
We multiply the upper equation with $(R^\prime_\mathrm{f}+r^\prime_\mathrm{g})$ and obtain
\be
R^\prime_\mathrm{f} V_6=r^\prime_\mathrm{g} V_7 + R^\prime_\mathrm{f} V_5,
\ee
which can be rewritten as
\be
y_2=\frac{V_7}{V_6-V_5}=\frac{R^\prime_\mathrm{f}}{r^\prime_\mathrm{g}}.
\label{eq:delta}
\ee
As per the parameters from Table~(\ref{tbl:values})
$R^\prime_\mathrm{f} \equiv R^\prime_\mathrm{F} = 10~\mathrm{k}\Omega$,
$r^\prime_\mathrm{g} \equiv R_\mathrm{G} = 100~\Omega$ and $y_2 \approx 100$.
Eq.~(\ref{eq:delta}) describes the amplification of a difference amplifier.
Buffer followed by a difference amplifier is called an instrumental amplifier.
For the instrumental amplifier considered by us $y_1 y_2 \approx 10100$ and in this way we are gradually entering in the terminology used in electronics.
\item In Fig.~\ref{Inverting amplifier} an amplifier, which can be considered as a difference amplifier with grounded (+) input is presented.
Then from Eq.~(\ref{eq:delta}) we obtain
\be
y_3=\frac{V_9}{V_8}=-\frac{R^\prime_\mathrm{f}}{r^\prime_\mathrm{g}}.
\ee
The substitution of parameters from Table~(\ref{tbl:values})
$R^\prime_\mathrm{f} \equiv R^\prime_\mathrm{F} = 10~\mathrm{k}\Omega$,
$r^\prime_\mathrm{g} \equiv R_\mathrm{G} = 100~\Omega$ gives $y_3 = 100$
and then the full gain of our circuit is
\be
Y=y_1 y_2 y_3 \approx -1.01 \times 10^6.
\ee

Until now we have studied static cases, for which we have direct voltages and currents or in other words, at zero frequency $f=0$.
In Fig.~\ref{fig:circuit}, however, we have capacitors $C_\mathrm{G}$ sequentially connected to the resistors  $R_\mathrm{G}$, and capacitors $C_\mathrm{F}$ and $C^\prime_\mathrm{F}$ parallely connected respectively to the resistors $R_\mathrm{F}$ and $R^\prime_\mathrm{F}$.
Despite this, the amplification of about million times is applicable in the frequency interval, for which
\be
\frac{1}{\omega C_\mathrm{G}} \ll R_\mathrm{G},
\ee
and in the meantime
\be
\omega C^\prime_\mathrm{F} \ll \frac{1}{R^\prime_\mathrm{F}},
\ee
where $\omega= 2 \pi f$ is the angular frequency.
The inequalities describe the negligible influence of the capacitors.
The large capacitor $C_\mathrm{G}$ should have negligible impedance in comparison with the resistor $R_\mathrm{G}$.
And in the meantime, the small capacitor $C^\prime_\mathrm{F}$ should have negligible conductivity in comparison with the conductivity of the resistor $R^\prime_\mathrm{F}$.
And finally, at a fixed resistor $R$ at the amplifier input, for the whole set of different input capacitors $C_\mathrm{i}$ the respective time constant $RC_\mathrm{i}$ should be in the optimal interval, in which the amplifier works
\be
R^\prime_\mathrm{F} C^\prime_\mathrm{F} \ll RC_\mathrm{i} \ll R_\mathrm{G} C_\mathrm{G}.
\ee
Because of
$R_\mathrm{G} C_\mathrm{G}/R^\prime_\mathrm{F} C^\prime_\mathrm{F} \approx 10^4 \gg 1$,
the demanding conditions are satisfied and the corrections to the simplified study at zero frequency are only few percent.
\item Let us consider and the non-linear element that ensures the measurement of the mean squared voltage of the amplified signal $U_X$.
In Fig.~\ref{Analog multiplier} a device is shown, for which the voltage at the output $U_W$ and the input voltages $U_X, U_Y$ and $U_Z$ are connected with the relation
\be
U_W=\frac{U_X U_Y}{U_\mathrm{m}}+U_Z,
\label{eq:mult}
\ee
where for the used in the set-up multiplier AD633 the constant is $U_\mathrm{m}=10$~V.
The current $I_W = U_W/(R_1+R_2)$, which goes down from the point W towards the ground creates a voltage at the point Z 
\be
U_Z=I_W R_2 = \frac{R_2}{R_1+R_2} U_W.
\ee
We substitute $U_Z$ in Eq.~(\ref{eq:mult}) and additionally considering that according to the connectivity from Fig.~\ref{Analog multiplier} $U_Y=U_X$, we obtain
\begin{align}
& U_W=\frac{U_X^2}{U_\mathrm{m}}+\frac{R_2}{R_1+R_2} U_W \quad \mbox{or} \nn \\
& U_W=\frac{U_X^2}{U_\mathrm{m}}\frac{R_1+R_2}{R_1}=
U_X^2/\tilde U, \quad \tilde U =U_\mathrm{m} \frac{R_1}{R_1+R_2}.
\label{eq:uw}
\end{align}
\item Finally, the circuit ends with averaging filter with large time constant
$R_\mathrm{av}C_\mathrm{av} = 15$~sec.
After this filter we observe the time averaged voltage $\left<U_W\right>$.
This averaged voltage through the averaging resistor $R_\mathrm{av}$ and the voltmeter with internal resistance $R_\mathrm{V}$ creates an averaged current 
\be
I_\mathrm{V}=\frac{\left < U_W \right >}{R_\mathrm{av}+R_\mathrm{V}}
\ee
and the averaged voltage measured by the voltmeter is
\be
U_\mathrm{V}=\frac{R_\mathrm{V}}{R_\mathrm{av}+R_\mathrm{V}} \left < U_W \right >.
\label{eq:uv}
\ee
In this way, combining the latter equation with Eq.~(\ref{eq:uw}) we obtain
\be
U_\mathrm{V}=\frac{\left < U_X(t)^2 \right >}{U_\mathrm{m}} \frac{R_1+R_2}{R_1}
\frac{R_\mathrm{V}}{R_\mathrm{av}+R_\mathrm{V}} =
\frac{\left < U_X(t)^2\right >}{U_0},
\label{eq:meas}
\ee
where for brevity we introduce the notation
\be
U_0= \frac{R_1}{R_1+R_2}\frac{R_\mathrm{av}+R_\mathrm{V}}{R_\mathrm{V}}U_\mathrm{m}.
\ee
The parameter with dimension voltage $U_0$ can be experimentally measured, if at the point X we apply with a potentiometer different direct voltages and measure the voltmeter voltage.

\end{enumerate}

Now we can synthesise the whole set-up operation.
The thermal voltage of the capacitor at the input $U(t)$ is amplified
\be
Y=y_1 y_2 y_3 = - \left ( \frac{2 R_\mathrm{F}}{r_\mathrm{_G}} + 1\right )
\frac{R^\prime_\mathrm{F}}{R_\mathrm{G}} \frac{R^\prime_\mathrm{F}}{R_\mathrm{G}}
\label{eq:final}
\ee
times $U_X(t)=YU(t)$, after which it is squared, averaged and measured with a voltmeter
\be
U_\mathrm{V}=\frac{\left < U(t)^2 \right >Y^2}{U_0}
= \frac{\left <U_X(t)^2\right >}{U^*},
\ee
where
\be
U^* \equiv \frac{U_\mathrm{m}}{Y^2}
\frac{R_1}{R_1+R_2} \frac{R_\mathrm{av}+R_\mathrm{V}}{R_\mathrm{V}}=\frac{U_0}{Y^2}.
\ee
So, if the equipartition theorem gives
\be
C \left < U(t)^2 \right > = k_\mathrm{_B} T,
\ee
the substitution in Eq.~(\ref{eq:final}) gives
\be
U_\mathrm{V}=\frac{k_\mathrm{_B} T}{CU^*} + \mathrm{const}.
\ee
The so obtained formula is used for the Boltzmann constant measurement in subproblems~23-25, where $y=U_\mathrm{V}$ and $x=T/CU^*$.

In conclusion, combining 6 simple problems on Ohm's law we obtained the complete theory of operation of our experimental set-up and the method for the processing of the obtained experimental data.

\section{EPO5}

This problem was given at the 5$^\mathrm{th}$ Experimental Physics Olympiad and more than 137 from the participants knew Ohm's law and managed not only to understand after the Olympiad but during the competition to do the elementary calculations.
Except for Ohm's law as mathematics, we used only the solution of a linear equation and elementary substitutions.
The areal of comprehensive high school students was between Ohrid and Astana, which is larger than the distance between Los Angeles and New York.
And it is not quite necessary to be a professional with expertise in the field of the analogue electronics to analyse circuits for which the voltages at the points (+) and (-) of some triangle are equal.
The high school students solve a series of 6 problems on Ohm's law with ease and they do not need a course in analogue electronics at all.
Moreover, the teachers that went through some similar course cannot remember something useful for the education and especially for the solution of the problem.
Inspired students as from Ohrid, as well as from Astana, took part in the Olympiad and if we have to compare with other continents, the distance between these two cities is larger than the distance between Los Angeles and New York.

\section{Advanced electronics applied to our setup}

And after all, at an engineering or at least university level, if we wish to reach 1\% accuracy, it is necessary to account for the frequency dependence emerging from the main equation of the operational amplifiers in frequency representation
\be
U_+-U_-=\varepsilon U_0, \quad \varepsilon(\omega) \approx \mathrm{j} \omega \tau, \quad
\mathrm{j}^2=-1, \quad \tau = \frac{1}{2 \pi f_0},
\label{eq:master}
\ee
where $f_0$ is the cut-off frequency of the operational amplifier.
The determination of $f_0^\mathrm{ADA4898} \approx 40$~MHz is described in Ref.~\onlinecite{Manhattan}.
A detailed calculation of the frequency dependency of the gain coefficients of the amplifiers is given in Ref.~\onlinecite{Detailed}, here we simply write down their values only
\begin{align}
& \Upsilon_1=\frac1{Y^{-1}(\omega)+\varepsilon(\omega)},
\quad Y(\omega)=\frac{Z_\mathrm{f}(\omega)}{z_\mathrm{g}(\omega)}+1, \quad
\Upsilon_1(0)=y_1=\frac{R_\mathrm{f}}{r_\mathrm{g}}+1, \quad
z_\mathrm{g}=r,
\quad \frac1{Z_\mathrm{f}}=\frac1{R_\mathrm{f}}+\mathrm{j}\omega C_\mathrm{f}, \\
& \Upsilon_2= \frac1{\Lambda(\omega)+\varepsilon(\omega)\left[\Lambda(\omega) +1\right]}, \quad
\Lambda(\omega)=\frac{z_\mathrm{g}(\omega)}{Z_\mathrm{f}^\prime(\omega)}, \quad
\Upsilon_2(0)=y_2=\frac{R_\mathrm{f}^\prime}{r_\mathrm{g}^\prime},
\quad z_\mathrm{g}=r_\mathrm{g}^\prime+\frac1{\mathrm{j}\omega C_\mathrm{g}},
\quad Z_\mathrm{f}^\prime=R_\mathrm{f}^\prime, \\
& \Upsilon_3=-\frac{1}{\mathcal{L}(\omega)+\varepsilon(\omega)\left[\mathcal{L}(\omega)+1\right]},
\quad \mathcal{L}(\omega)=\frac{z_\mathrm{g}(\omega)}{Z_\mathrm{f}^{\prime\prime}(\omega)},
\quad \Upsilon_3(0)=y_3=-\frac{R_\mathrm{f}^\prime}{r_\mathrm{g}^\prime},
\quad \frac{1}{Z_\mathrm{f}^{\prime\prime}}=\frac{1}{R_\mathrm{f}^\prime}+
\mathrm{j} \omega C_\mathrm{f}^\prime, 
\end{align}
and for the whole frequency dependent gain coefficient of the amplifier we obtain the product of the separate independent amplifiers
\be
\Upsilon(\omega)=\Upsilon_1 \Upsilon_2 \Upsilon_3.
\ee
After calculating the pass bandwidth of the amplifier, which is an important notion in electronics, we derive the substitution rule
\be
\frac{1}{U^*}=\frac{1}{U_0}Y^2 \rightarrow \frac{1}{U_0} Y^2 Z,
\ee
where the correction multiplier $Z$ is given with the integral describing the amplifier pass bandwidth and the filter at its input
\be
Z=1+\epsilon
=\dfrac{\int_0^\infty \frac{\left |\Upsilon(\omega)\right |^2}{1+(\omega RC)^2} \mathrm{d}\omega}
{ \int_0^\infty \frac{Y^2}{1+(\omega RC)^2} \mathrm{d}\omega}
=\frac{2}{\pi} \dfrac{RC}{Y^2} \int_0^\infty \frac{\left |\Upsilon(\omega)\right |^2}{1+(\omega RC)^2} \mathrm{d}\omega.
\ee
The integral in the denominator gives the pass bandwidth in case of amplifier operation up to high frequencies and the capacitors, which stop the floating of the zero and the self-excitation are not mounted.
The numerical calculation of the correction gives $\epsilon \approx 6\%$, with which all systematic errors for our used set-up are accounted for.
In this formula the capacitances $C_n$ from the examined set should be substituted.

\section{Lock-in voltmeter hidden in the experimental set-up}

Let us now take a look at the element named S\textunderscore M in Figs.~\ref{PCB} and \ref{fig:circuit}.
In the circuit in Fig.~\ref{fig:circuit} S\textunderscore M is a switch, whose initial position is closed, that equals the potentials of both ``IN'' and ``NC'' inputs.
Or in other words, $U_X=U_Y$, as we have already considered in the operation of the non-linear element.

For the Olympiad problems the switch S\textunderscore M is not necessary and hence it has not been placed on the printed circuit board in Fig.~\ref{PCB}.
Instead, both switch contacts are soldered together, which in practice means that the switch is permanently closed or short-circuited (i.e. missing).
The joint connecting both switch contacts can be easily removed by cutting with diagonal pliers (wire cutters) or unsoldering it and in this way the switch is opened and now $U_X \neq U_Y$, which means that the multiplier multiplies two different signals.

Let us now consider a different element in principle.
A weak signal $U_s(t)=U_s \cos(\omega t)$ is applied at the amplifier input, which is amplified $Y$ times, so that at the multiplier $X$ input voltage $U_X(t)=Y U_s(t)$ is applied.
In case of a removed connection between $X$ and $Y$ inputs, i.e. open switch S\textunderscore M in Fig.~\ref{PCB}, at the ``NC'' labeled input of the board a known signal $U_r(t)=U_r\cos(\omega t)$ is applied.
This signal goes to the $Y$ multiplier input $U_Y(t)=U_s(t).$

Then the multiplied signal according to Eq.~(\ref{eq:mult}) will be
\be
U_W=\frac12 \frac{YU_s(t) U_r(t) \cos^2(\omega t)}{U_\mathrm{m}}+U_Z.
\ee
We have already expressed $U_Z$ by the resistances $R_1$ and $R_2$, and that is why directly using Eqs.~(\ref{eq:uv}) and (\ref{eq:meas}) and accounting that $ \left < \cos^2(\omega t) \right > = 1/2$, for the time averaged voltaged measured by the voltmeter we obtain
\be
U_V=\frac{Y U_r}{2 U_0}U_s .
\ee
In this way, at the circuit output the voltmeter measures direct voltage $U_V$, in which the studied by us periodic signal with amplitude $U_s$ is present.
The signal $U_r$ that we have applied to the $Y$ input is called carrier or reference signal and a device whose operation we have just described is also called a lock-in voltmeter (amplifier).

Lock-in amplifiers are used for small signals measurements even when through an oscilloscope the amplified signal is lost in the background of the external noise.
In this synchronous measurement of two sinusoidal signals with common frequency, the measured signal can be separated from the noise with spectral density
$(\mathcal{E}^2)_f$, if
\be
U_s>\sqrt{(\mathcal{E}^2)_f \,\Delta f_\mathrm{av}},
\quad \Delta f_\mathrm{av}=\frac1{2\pi R_\mathrm{av}C_\mathrm{av}},
\ee
i.e. at large enough time averaging
$\tau_\mathrm{av}=R_\mathrm{av}C_\mathrm{av}$.
Still the noise should not overload the amplifier
\be
U_m>\sqrt{Y^2\, (\mathcal{E}^2)_f\,B},
\quad B= \frac{f_0}{y_1}
\ee
and this sets the upper limit of the possible gain coefficient
\be
Y<\frac{U_m}{\sqrt{(\mathcal{E}^2)_f\,B}}.
\ee

\section{EPO5 feedback in English}

\begin{itemize}

\item Adaptation of a computer translation from Greek.

Experiences of the 5th Balkan Olympiad in Experimental Physics ``The day of the electron"

Thank you for the opportunity you gave me to attend the Olympics by gathering valuable information, knowledge of how this institution is being conducted and exchanging views with other countries' science teachers on science education.
I would like to congratulate all students from the Greek team who participated in the competition by writing history as the first Greek mission.
The Balkan Olympiad was organized by the Bulgarian Physicists' Section, with the collaboration of the Physics Faculty of the Sofia University, and was held at the St.Clement University of Ohrid University in the Physics Faculty.
Over the years the Olympiad has been established as a tradition for extracurricular education in physics.
The aim of this year's Olympiad was to present an experimental problem by exploring the properties of an electron.
The authors of the works used the experiments described in the textbooks manuals as part of the curriculum.
On the first day all participants and their escorts gathered in the event hall.
150 pupils from the Balkans and Russia participated.
The opening speech was begun by the dean of the Physics Faculty (prof. Dreischuh), the contest manager and then the speakers from different countries.
The enthusiasm was too great.
Then, the students were divided into sections of about 15 people of different nationalities and left for the university classes.
They had at their disposal 4 hours to experiment experimentally with the experimental exercises of the experiment that 111 years ago, the great physicist Einstein, a pioneering method for measuring the Boltzmann constant by studying the average thermal energy of a capacitor.
Surprisingly, to date, this experiment has never been done, with the result that the participants in the competition, following the appropriate processing of their experiment and methodology by the university's university team, and the development of technology over the last 100 years that allowed it to adapt of the scientific problem at school level, were the first to apply it in practice.
At first, students were provided with the necessary materials to carry out the experiment, such as electronic boards, batteries and terminals.
The pupils had to be familiar with the use of measuring instruments (multimeters) that they brought with them.
As the students competed, the teachers gathered in another room and conducted the experiment at the same time.
Prior to the experiment, I found that the brochures were not translated into the Greek language, nor was there a group of explanations of the issues for our country, which made it difficult for the students of the Greek team to fully understand the exercises. The organization of the contest was impeccable.
The next day we went back to the University that followed the award ceremony bringing the student from Kazakhstan to the contest. Sofia greeted us with a dreamy snowy landscape.

\item 
The Olympiad was truly great and I had real fun conducting the experiment, which I plane to conduct with my classmates one day at school. I have a few recommendations for the next Olympiad as well. First of all, it would be good to be are informed about the instruments that we will have to use during the experiment. I personally had a problem with the potentiometer since this was the first time I ever got to connect it to the circuit and I had to spend time in order to find out how. This resulted in spending a big amount of time and then not having enough to design a nice graphical representation of the experimental data. Besides that, it would also be good to have at least one thermometer in every classroom in order to determine the room temperature, because guessing it makes the experiment inaccurate. Other than that, I believe the organization of the Olympiad was great. The experimental setup is truly amazing and I would like to say a big thank you for offering it to us for free. I will try to use it in the best way possible. It was an honour to have been part of this Olympiad.

\subsection{Answers to the question: Have you repeated the experiment at home? What value did you get for Boltzmann's constant?}

\item Yes, we repeated it in the physics class.
\item Yes, 15\% error.
\item $1.29\times 10^{-23}$J/K.
\item I have not repeated the experiment at home yet ,but I am planning to do so in a week during the Christmas holidays , since I am going to have a lot of free time . Nevertheless , I have actually kept the measurements I did during the EPO and I analyzed them using Microsoft Office Excel. The value I got was k=1.182*10$^{-23}$~ J/K  (at a temperature of 17 degrees Celsius) which is not very close ( there is a 15\% error compared to the known value , k=1.38*10$^{-23}$~J/K). I concluded that I should have measured the voltage U2,n ( the one measured in task 18) more precisely, in mV.   

\section{The result of the absolute champion}

\begin{figure}[h]
\includegraphics[scale=0.36]{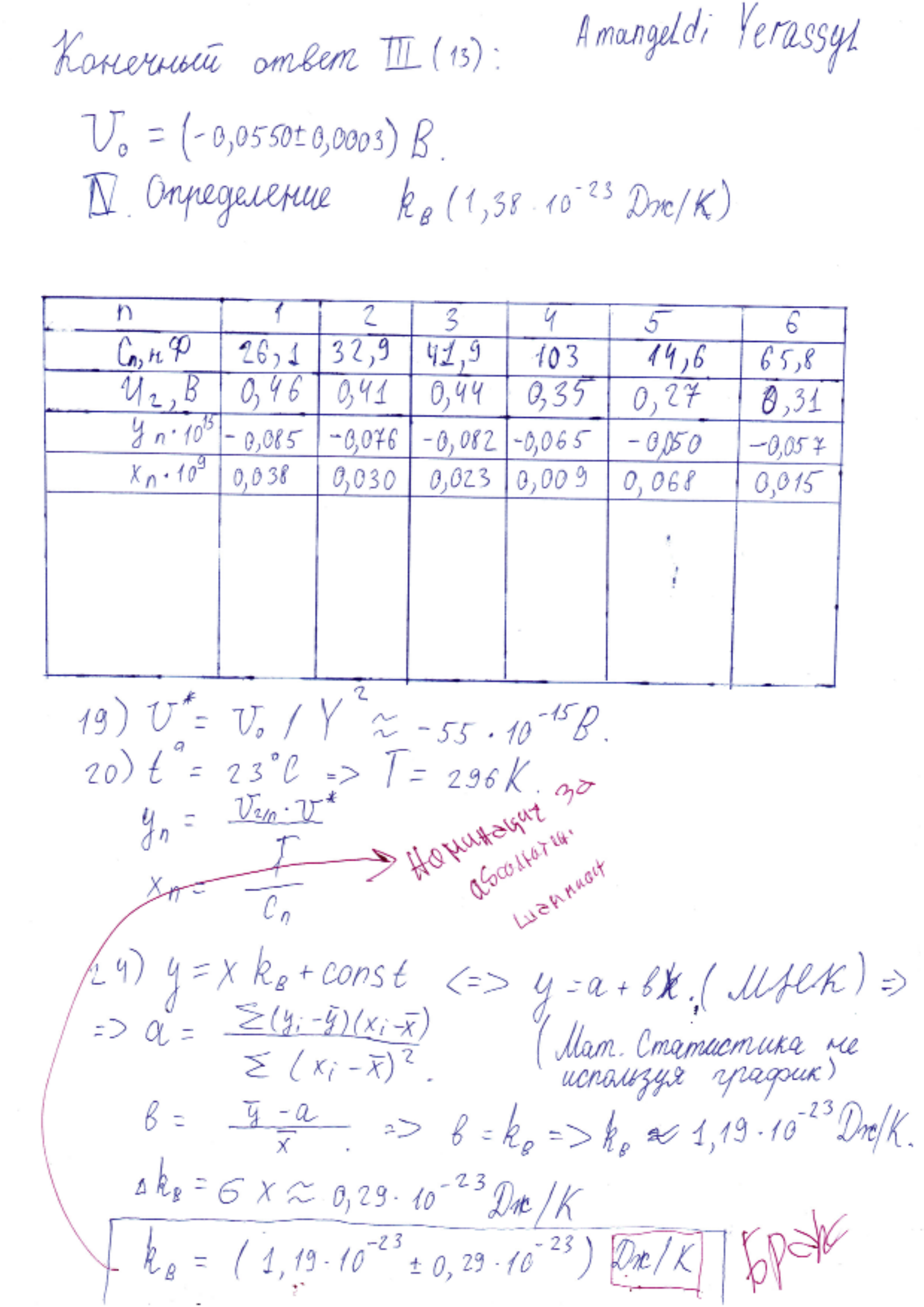}
\end{figure}

\end{itemize}

\newpage
\begin{center}
\large \textbf{Измерване на константата на Болцман по Айнщайн. Задача на 5-тата Oлимпиада по експериментална физика. София 9 декември 2017}
\end{center}

\begin{center}
\textbf{Тодор~М.~Мишонов$^1$, Емил~Г.~Петков$^1$, Aлександър~А.~Стефанов$^1$, Александър~П.~Петков$^1$, Иглика~М.~Димитрова$^2$, Стоян~Г.~Манолев$^3$, Симона И. Илиева$^4$, Алберт~М.~Варонов$^5$}
\end{center}

\begin{center}
\raggedright
$^1$Софийски университет Св. Климент Охридски, \\
Физически факултет, София 1164, бул. Джеймс Баучер 5

$^2$Химикотехнологичен и металургичен университет, \\
Факултет по химични технологии, к-ра Физикохимия,
София 1156, бул. Климент Охридски 8

$^3$СУ Гоце Делчев, 
ул. Първомайска 3, MKD-2460 Валандово, Република Македония

$^4$Софийски университет Св. Климент Охридски, \\
Физически факултет, к-ра Атомна Физика,
София 1164, бул. Джеймс Баучер 5

$^1$Софийски университет Св. Климент Охридски, \\
Физически факултет, к-ра Теоретична физика, София 1164, бул. Джеймс Баучер 5
\end{center}

\begin{center}
\indent \textbf{Резюме.}
\raggedright
Описани са няколко последователни експеримента със специално конструирана постановка. 
Изпълнението на последователните експериментални задачи дава възможност за определяне на константата на Болцман $k_\mathrm{_B}$. 
Изследват се флуктуациите на напрежението U(t) на серия от кондензатори 
свързани успоредно с постоянно съпротивление. 
Напрежението се усилва 1~милион пъти $Y=10^6$. 
Усиленото напрежение YU(t)  се подава на устройство,
което отчита средния по време квадрат на напрежението
$U_2=\left<(Y U(t))^2\right>/U_0$.
Напрежението $U_2$ се измерва с мултиметър. 
Серия от измервания дава възможност за определяне на константата на Болцман от
теоремата за равноразпределението $C\left<U^2\right>=k_\mathrm{_B}T$.
Средната кинетична енергия на атомното движение на един газ 
$\frac12 m \mathbf{v}^2=\frac32 k_\mathrm{_B}T$
е най-известния пример от ученическата физика.
За определяне на константата на постановката $U_0$
е дадена серия от задачи свързани със закона на Ом, които са 
адресирани към най-големите ученици.
За най-малките ученици основната задача е да се изследва аналоговото повдигане на квадрат.
Работите на учениците се оценяват в 4 възрастови групи S, M, L  и XL.
Последните подусловия са предназначени за студенти (категория XL) и
са свързани с теоретичното изследване на постановката като инженерно съоръжение.
Тази задача е дадена на Петата Олимпиада по Експериментална Физика ``Ден на Електрона'' на 9 декември 2017 г., София, организирана от Софийския клон на Съюза на физиците в България със съдействието на Регионалното дружество на физиците в Струмица, Р.~Македония и Физически факултет на СУ.
\end{center}
\captionsetup{labelfont={normalsize},textfont={small},justification=centerlast}

\section{Увод}

От самото си начало Олимпиадата по експерименталната физика (ОЕФ) е световно известна;
задачите на всички Олимпиади са публикувани в Интернет~\cite{EPO1,EPO2,EPO3,EPO4}, а от самото начало участниците бяха 120.
В последните години са участвали ученици от 7 страни и разстоянието между най-отдалечените градове е повече от 4~Mm.

Нека опишем главните различия между ОЕФ и други подобни състезания.
\begin{itemize}
\item Всеки участник на ОЕФ получава като подарък от организаторите постановката, с която е работил.
Така след Олимпиадата, даже лошо представил се участник, ще може да повтори експеримента и да достигне нивото на шампиона.
По този начин Олимпиадата пряко повлиява върху нивото на преподаване в целия свят.
След приключване на учебната година, постановката остава в училището, където е учил участника.
\item Всяка от задачите е авторска и е свързана с фундаментална физика или разбирането на работата на технически патент.
\item Олимпийската идея се реализира в ОЕФ в първоначалния си вид и всеки желаещ от целия свят може да участва.
Няма ограничение на броя участници.
От друга страна, приликата с другите Олимпиади е, че задачите са пряка илюстрация на учебния материал и заедно с други подобни състезания смекчава деградацията на средното образование, което е световна тенденция.
\item Една и съща експериментална постановка се дава за всички участници, но задачите са различни за различните възрастови групи, така както водата в басейна е еднакво мокра за всички  възрастови групи при едно състезание по плуване.
\end{itemize}

Нека изброим накратко задачите от отминалите 4 ОЕФ.
Постановката на първата олимпиада ОЕФ1 бе една ученическа версия на
американския патент за принципа на самонулиране на постоянно токови усилватели~\cite{EPO1}.
Задачата на втората ОЕФ2~\cite{EPO2} бе да се измери константата на Планк като се използва 
дифракция от компакт диск на светлина от светодиоди .
Една съвременна реализация на принадлежащия на НАСА патент 
за използване на отрицателното съпротивление за генериране на електрични трептения 
бе постановката на ОЕФ3~\cite{EPO3}.
Четвъртата ОЕФ~\cite{EPO4} бе посветена на фундаменталната физика
-- да се определи скоростта на светлината чрез измерване на електрични и магнитни сили.
Иновативния елемент бе приложението на теорията на катастрофите
към анализа на устойчивостта на махала.
Вече установената традиция за стила на олимпиадата 
е баланса между съвременни работещи технически изобретения и фундаменталната физика.

И тази 5-та Олимпиада (ОЕФ5) следва установената традиция.
В учебните програми на всички страни се споменава, че топлината е свързана с движението на атомите.
Но даже сравнително прост пример на Максвелово разпределение на молекулите по скорости не се илюстрира с експеримент даже в много добри университети.
Това е защото вакуумната техника е много скъпа и изисква професионална работа.
От друга страна, електронните измервания са устойчиви (\textit{дуракоустойчиви, foolproof}) и хиляда пъти по-евтини.
Затова, ако искаме да изследваме енергията на топлинните флуктуации, която се описва от температурата и константата на Болцман~\cite{McCombie:71}
\be
m \left< v_x^2 \right>= k_\mathrm{_B} T= C \left <U^2 \right >,
\nn
\ee
изследването на средноквадратичното напрежение $\left <U^2 \right >$ е хиляди пъти по-евтино от измерването на средноквадратичната скорост $\left< v_x^2 \right>$.
Затова, за методични цели при известен капацитет определянето на константата на Болцман чрез електрични измервания е единствено възможния метод за средното образование.
Като фундаментална физика, тази идея бе изказана от Алберт Айнщайн, а сега 111 след това методът е експериментална задача за ученици, а постановката е изготвена в 200 екземпляра.

\section{Задача на настоящата ОЕФ5} 
Преди 111 години Алберт Айщайн~\cite{Einstein:07} предложи метод
за измерване на константата на Болцман 
$k_\mathrm{_B}$ чрез изследване на средната топлинна енергия 
$C\left<U^2\right>=k_\mathrm{_B} T$
на кондензатор с капацитет C. 
Странното е, че досега този експеримент не е извършен и вие сте първите, които ще могат да го направят след като нашият колектив реализира тази идея.
Развитието на техниката през изминалото столетие направи възможно този научен проблем да бъде адаптиран за нивото на ученическа задача.
Започваме с подусловията, които са решими и от най-малките ученици
и трудността постепенно ще нараства като работите докъдето можете да стигнете.
Последните подусловия са адресирани към студентите.

Теми за по-нататъшна работа.
Участниците от олимпиадата получават от организаторите експерименталната постановка, с която са работили, като една от най-важните цели на Олимпиадата е внимателното повтаряне на експеримента до детайлното измерване на константата на Болцман и пълното разбиране на теорията. 
Решението на задачата ще бъде публикувано на сървъра на библиотеката на Корнелския университет, 
 а няколко детайла от сканирани работи на участници в Олимпиадата ще бъдат публикувани на сайта на Софийски клон на Съюза на физиците в България.
Очакваме участниците в олимпиадата, след като прочетат решението, да извършат всички описани в него експерименти и да получат константата на Болцман $k_\mathrm{_B}$. 
Това първо определяне на фундаментална константа хвърля върху нас отблясъка на една велика епоха в развитието на физиката, когато квантите и атомите са били предмет на забавление на момчета, създали съвременната физика.

Подобни експериментални задачи се дават само в
най-добрите университети~\cite{Wash:12, MIT:13} и който има достъп до описанието на постановките, нека сравни точността, която се получава при нашия ученически експеримент.  

Теоретичните подусловия са едно добро упражнение по закона на Ом, което дава възможност за цялостното разбиране на работата на постановката и тези задачи са едно добро начало за разбиране на електрониката, която стои в основата на много физични прибори.
Експерименталните постановки са подарък от организаторите на олимпиадата за участниците, които могат да ги ползват и да правят демонстрации в часовете по физика. След края на учебната година, тези постановки трябва да бъдат предадени на кабинета по физика в училището, където участниците учат.

\section{Начални лесни задачи. S}
\begin{enumerate}
\item Измерете напреженията на 4-те батерии от 9~V 
и с максимална точност на батерията от 1.5~V.
\item Поставете батерията от 1.5~V в държача, свържете я потенциометрично 
и измерете интервала от напрежения, които постигате когато въртите оста на потенциометъра.
Това ще бъде източника за напрежение за следващите задачи.
Обръщането на полярността променя знака на напрежението.
\item Свържете батериите от 9~V със съединителните клипсове, като закопчаете капачките.

\section{Аналогово повдигане на квадрат. M}
\item \textbf{Внимание! От този момент нататък вече има възможност да изгорите интегралните схеми на постановката, ако свържете батериите неправилно.}
Ориентирайте постановката, с която работите така, че двете крайни жички да са отдясно,
а етикета с надпис ``COM'' да бъде долу вдясно. Виж схемата на Фиг.~\ref{PCB}.
\item Внимателно включете рейката с батериите от 9~V 
към десните 3 рейки от дънната платка;
етикет към етикет, лист към лист.
Работете внимателно - ако сбъркате полярността ще изгорите интегралните схеми.
\item Свържете потенциометрично свързаната батерия  
с входната жичка на постановката горе по средата с надпис ``IN'', а другия електрод на потенциометъра 
към жичката, която излиза от ``земята'' на схемата с надпис ``COM''.
\item Успоредно на потенциометъра и ``IN''--``COM'' входовете на постановката
включете първия волтметър, който показва напрежение $U_1.$
\item При правилно свързване на схемата, когато въртите оста на потенциометъра,
напрежението $U_1$ трябва да се изменя приблизително между 
нулата и напрежението на батерията от 1.5~V.
\item Вторият волтметър, който показва напрежение $U_2$, включете между 
електрода на изхода на постановката ``OUT'' и електрода на общата точка ``COM''.
Така $U_2$ е напрежението между точките ``OUT''--``COM''.
\item Проверете дали ``COM'' електродите на двата мултиметра и постановката са свързани.
\item Въртете оста на потенциометъра, изчакайте 1 минута и запишете напреженията
$U_1$ и $U_2$. 
Обърнете полярността на източника на напрежение и повторете измерванията на входното напрежение $U_1$ и изходното напрежение $U_2$. 
Резултатите подредете в таблица със стълбове: 
номер на измерването $i$, $U_1$ и $U_2$. 
\item Резултатите представете графично в равнината $U_1$ по абсциса (хоризонтала), $U_2$ по ордината (вертикала).
\item Към таблицата добавете допълнителна колона $(U_1)^2$ и представете резултатите
графично в равнината $(U_1)^2$ абсциса и $U_2$ ордината.
Начертайте права линия, която най-добре преминава в близост до експерименталните точки.
Тази приближаваща (апроксимираща) права се описва с уравнението
$U_2=(U_1)^2/U_0 + \mathrm{const}.$
Върху правата изберете 2 точки, измерете разликите по абсциса $\Delta(U_1^2)$,
по ордината $\Delta(U_2)$, и по наклона определете параметъра с размерност напрежение
$U_0=\Delta(U_1^2)/\Delta(U_2)$.
Този параметър на аналогово умножаващата схема е съществен за определянето на константата на Болцман $k_\mathrm{_B}$ и това е описано в подусловията от следващия раздел.

\section{Флуктоскопия на напрежението. Определяне на константата на Болцман. L}

\begin{figure}
\includegraphics[scale=1.0]{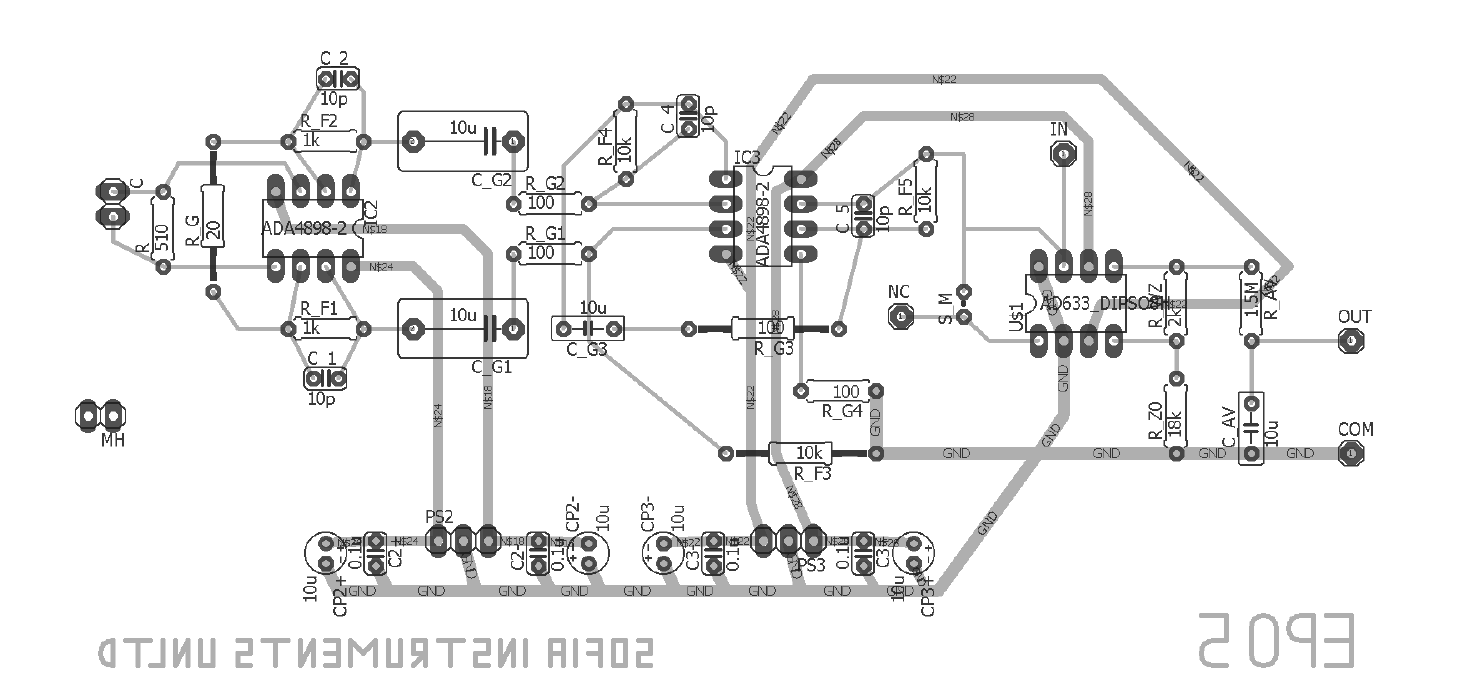}
\caption{PCB (Схема на печатната платка)}
\label{bPCB}
\end{figure}

\item Откачете потенциометъра и волтметъра, който измерва ``IN''--``COM'' напрежението $U_1$. 

\item Внимателно включете и втория източник за напрежение с две батерии от 
9~V. Етикет към етикет, оранжево листче към оранжево листче, 
иначе интегралната схема ще изгори. 

\item Сега предстои най-опасният момент за Интегралните схеми (ADA4898-2).
Трябва и двете интегрални схеми да се монтират към постановката.
Отново: зелено листче към зелено листче, етикет към етикет.
При грешка изгарят.

\item Вие разполагате с набор от кондензатори запоени на рейки. 
Капацитетът $C$ на всеки кондензатор в нанофаради ($\mathrm{1\;nF=10^{-9}\;F}$) 
е написан върху залепения на рейката етикет. 
Може да подредите кондензаторите по големина и да запишете
стойността на капацитета им $C_n$ в таблица; 
индексът $n$ показва поредния им номер.

\item 
Сега следва най-важното измерване -- изследването на топлинните флуктуации на напрежението на кондензатора. Дръжте жичката, свързана с етикета ``NC'' на платката, далеч от източниците на напрежение.
Последователно закачате всеки един от кондензаторите на рейката 
в левия край на постановката.
Изчаквате 2 минути и записвате напрежението $U_2$, което показва волтметърът
включен между ``OUT'' и ``COM'' електродите на постановката.
Резултатите подреждате в таблица със стълбове $n$, $C_n$ и $U_2$.

\item Флуктуиращото напрежение $U$ на кондензатора $C$ се усилва 
милион пъти, по-точно коефициентът на усилване е 
$Y=1.01\times 10^6,$ сравни с усилвателя на братя Хабихт.~\cite{Habicht:10} 
Пресметнете важния за постановката параметър 
с размерност напрежение 
$U^*=U_0/Y^2.$
Не се плашете от степените, това е едно много малко напрежение 
от порядъка на пико-волт ($\mathrm{1\;pV=10^{-12}\;V}$).

\item Преценете колко е температурата в аудиторията и я запишете в Келвини (0~K$\; \approx -273^\circ$C).

\item Допълнете таблицата с две нови колони:
$y_n=U_{2,n} U^*/T$ и $x_n=1/C_n$;
това са най-важните експериментални данни!

\item Резултатите от таблицата представете графично в равнината $x$-$y$.
Отделете няколко минути за да обмислите мащаба.

\item При правилна работа на постановката отделните експериментални точки 
са в близост до една права.
Начертайте правата $y=k_\mathrm{_B}\,x + \mathrm{const}$, която по ваша преценка най-добре преминава в близост до точките. 
Тази стандартна математична процедура се нарича линейна регресия 
и е заложена в много калкулатори, ако искате да получите решението числено.

\item Върху правата изберете две точки и определете разликата 
в техните координати по вертикала $\Delta y$ и хоризонтала $\Delta x$.
Запишете тези разлики.

\item Пресметнете наклона на правата $k_\mathrm{_B}=\Delta y/\Delta x$
и начертайте рамка около вашия резултат.
Поздравления, Вие измерихте фундаменталната константа на Болцман $k_\mathrm{_B}$ и дори правилно определеният порядък е голям успех.

\section{Теоретични задачи за закона на Ом отнасящи се за постановката. XL}

Олимпиадата е по експериментална физика и задачите от тази секция дават малко точки, които могат да окажат незначително влияние върху класирането. 
Резултатите от решението на тези подусловия могат да се използват 
само за разграничаване на работи, в които експеримента е изпълнен еднакво добре
или пък за ненулев резултат на работи, в които участникът 
е изгорил интегралните схеми, нарушавайки инструкциите за монтирането.

На схемите показани на Фиг.~\ref{bNon-inverting amplifier}, \ref{bBuffer}, \ref{bDifferential amplifier} и \ref{bInverting amplifier}
има нарисувани триъгълници.
На вертикалната страна отляво влизат проводници 
и входовете са маркирани с (+) и (-),
а от десният връх излиза проводник и понякога се добавя маркировка (0).
Устройството работи така, че от изхода (0) излиза такъв ток $I_0$, 
че напреженията на входовете (+) и (-) се изравняват
$U_+=U_-$.
Токовете, които влизат във входовете (+) и (-) са пренебрежими.
Такова устройство се нарича операционен усилвател.

\item Като използвате закона на Ом покажете, 
че отношението на напреженията и съпротивленията са 
свързани с уравнението $y_1=V_2/V_1=1+R_f/r_g.$ 
Виж Фиг.~\ref{bNon-inverting amplifier}.

\item Аналогично за симетричния вариант на същата схема
$y_1=(V_2-V_4)/(V_1-V_3)=R_f/r_g+1$. Виж Фиг.~\ref{bBuffer}.

\item $y_2=V_7/(V_5-V_6)=-R_f/r_g$. Фиг.~\ref{bDifferential amplifier}.

\item $y_3=V_9/V_8 =-R_f/r_g$. Фиг.~\ref{bInverting amplifier}.

\item Ако голям кондензатор $C_g$ е свързан последователно с резистор $r_g$ 
покажете, че при честоти за които $\omega r_g C_g \gg 1$,
влиянието на кондензатора е пренебрежимо?

\item При какви честоти е пренебрежимо влиянието на малък кондензатор $C_f$ успоредно свързан с голям резистор $R_f$? 

\item Ако 3-те усилвателя които разгледахме са свързани последователно
общото им усилване е $Y = y_1 y_2 y_3.$ 
Oценете честотния интервал $(f_\mathrm{l},\,f_\mathrm{h})$
при който коефициента на усилване $Y$ е приблизително постоянен, 
като използвате числените стойности на параметрите 
от таблица~\ref{tbl:bvalues}; $f=\omega/2\pi$

\item На схемата от Фиг.~\ref{bAnalog multiplier} напрежението $U_W=U_X U_Y/U_m+U_Z$, 
като $U_X=U_Y$, а константата $U_m=10\;\mathrm{V}$ за използвания AD633.
Изразете $U_Z$ чрез $U_W$ и покажете, че $U_W=U_X^2/\tilde{U},$
където $\tilde{U}=U_m\,R_1/(R_1+R2).$

\item Покажете, че $U_2=R_\mathrm{V} \left<U_W\right>/(R_\mathrm{av}+R_\mathrm{V})$,
където $R_\mathrm{V}$ най-често $1\;\mathrm{M}\Omega$ е вътрешното съпротивление на омметъра,
а скобите на $\left<U_W\right>$ означават осредняване по време.

\item Покажете, че $U_2=U_X^2/U_0,$ където $U_0=\tilde U (R_\mathrm{av}+R_\mathrm{V})/R_\mathrm{V}$.

\item Накрая основното уравнение за нашата постановка
$U_2=\left< U^2\right>/U^{*}$,
където $U$ е напрежението на кондензатора на входа на прибора,
а $$\frac1{U_*} \equiv \frac{Y^2}{U_0} =
\frac{Y^2}{U_m}\frac{R_1+R_2}{R_1}\frac{R_\mathrm{V}}{R_\mathrm{av}+R_\mathrm{V}}.
$$

\item Пресметнете параметъра $U_*$ като използвате параметрите от Таблица~\ref{tbl:bvalues}.

\begin{figure*}[h]
\includegraphics[scale=0.25]{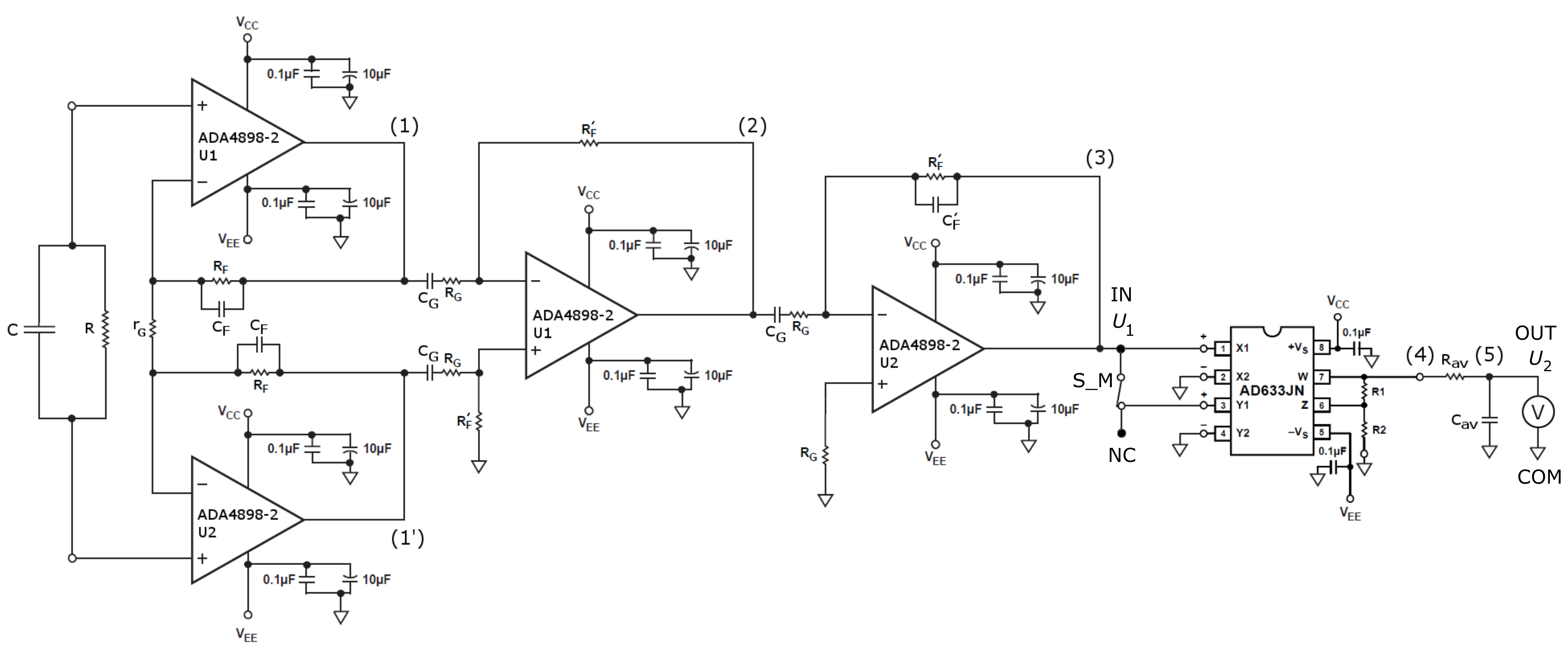}
\caption{Подробна схема на постановката.}
\label{fig:bcircuit}
\end{figure*}

\begin{center}
\begin{table}[h]
\begin{tabular}{| c | r |}
		\hline
		&  \\ [-1em]
		Елемент от схемата  & Стойност  \\ \tableline
			&  \\ [-1em]
			$R$ & 510~$\Omega$ \\
			$r_\mathrm{_G}$ & 20~$\Omega$ \\
			$R_\mathrm{F}$ &  1~k$\Omega$  \\
			$C_\mathrm{F}$ &  10~pF  \\ 
			$C_\mathrm{G}$ & 10~$\mu$F \\
			$R_\mathrm{G}$ &  100~$\Omega$  \\ 
			$R_\mathrm{F}^\prime$ & 10~k$\Omega$ \\
			$C_\mathrm{F}^\prime$ & 10~pF \\
			$R_1$ &  2~k$\Omega$  \\ 
			$R_2$ & 18~k$\Omega$  \\
			$R_\mathrm{av}$ & 1.5~M$\Omega$ \\
			$C_\mathrm{av}$ & 10~$\mu$F \\
			$R_\mathrm{_V}$ & $\approx 1~\mathrm{M} \Omega$ \\
			$V_\mathrm{CC}$ & +9~V \\
			$V_\mathrm{EE}$ & -9~V \\
\tableline
\end{tabular}
	\caption{Таблица на числените стойности на елементите от схемата показана на 
	Фиг.~\ref{fig:bcircuit}.}
	\label{tbl:bvalues}
\end{table}
\end{center}

\end{enumerate}

\begin{figure}[h]
\begin{minipage}[t]{0.31\linewidth}
\includegraphics[scale=0.28]{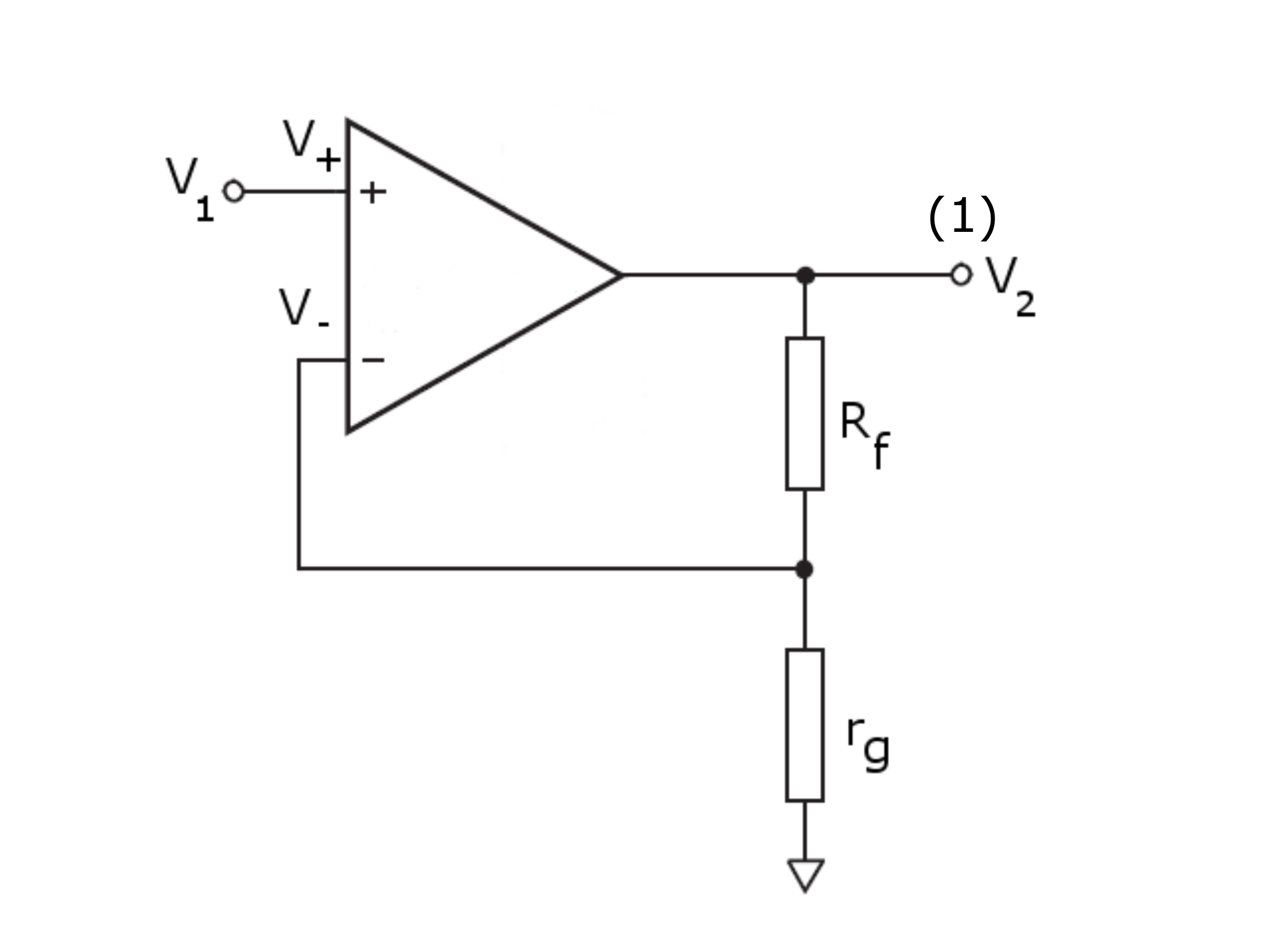}
\caption{Неинвертиращ усилвател}
\label{bNon-inverting amplifier}
\end{minipage}
\begin{minipage}[t]{0.31\linewidth}
\includegraphics[scale=0.28]{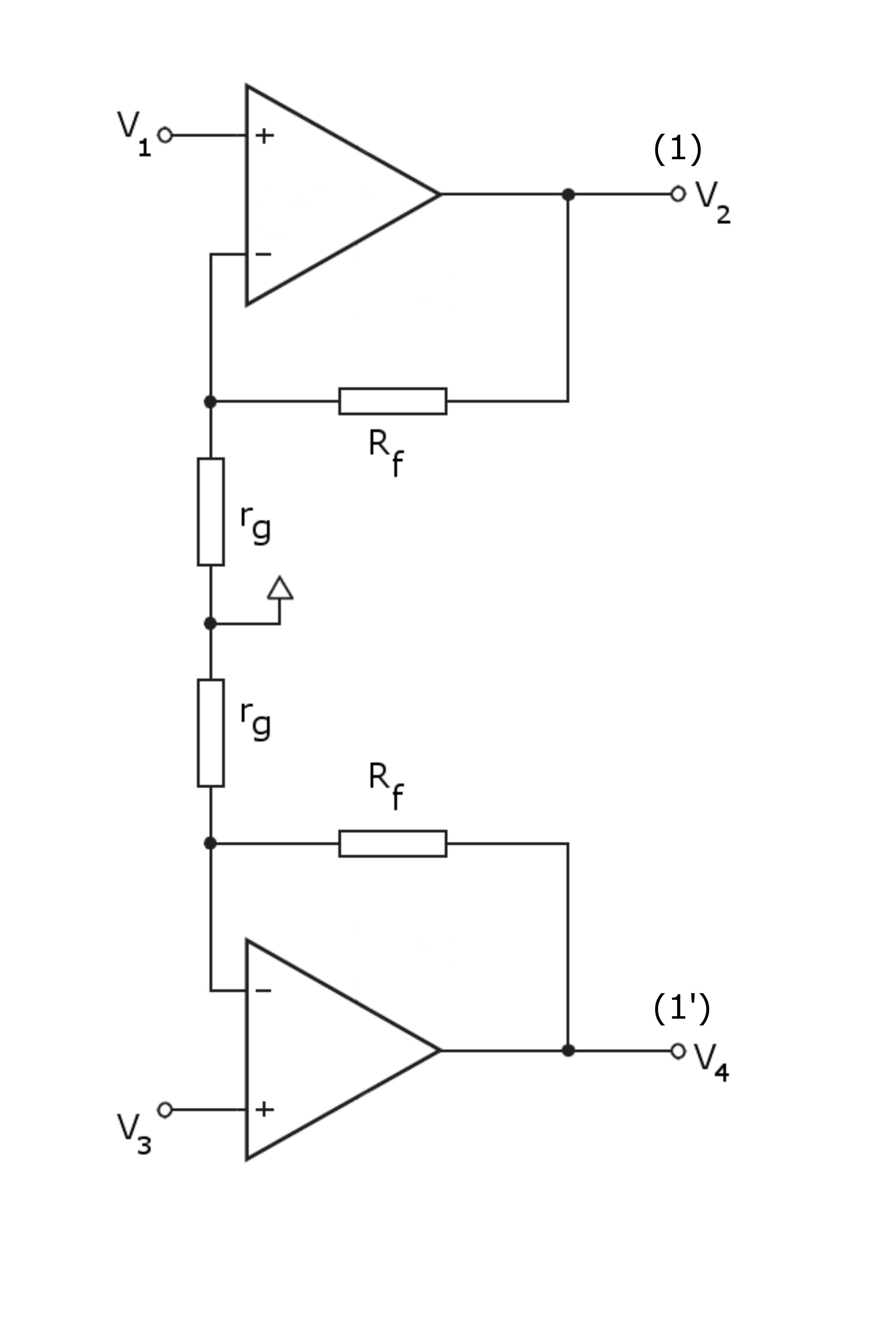}
\caption{Буфер}
\label{bBuffer}
\end{minipage}
\begin{minipage}[t]{0.36\linewidth}
\includegraphics[scale=0.28]{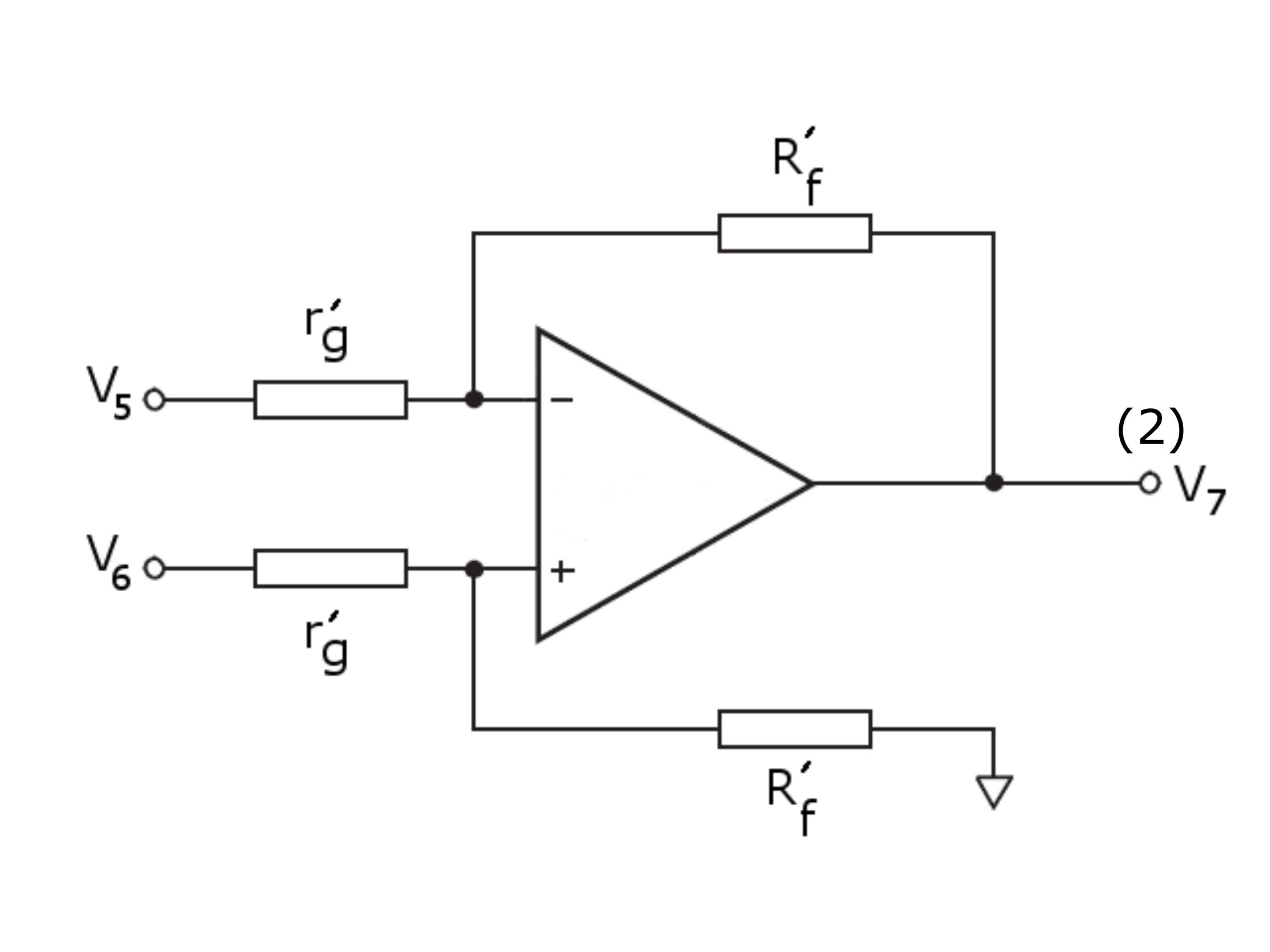}
\caption{Разликов усилвател}
\label{bDifferential amplifier}
\end{minipage}
\begin{minipage}[c]{0.4\linewidth}
\includegraphics[scale=0.28]{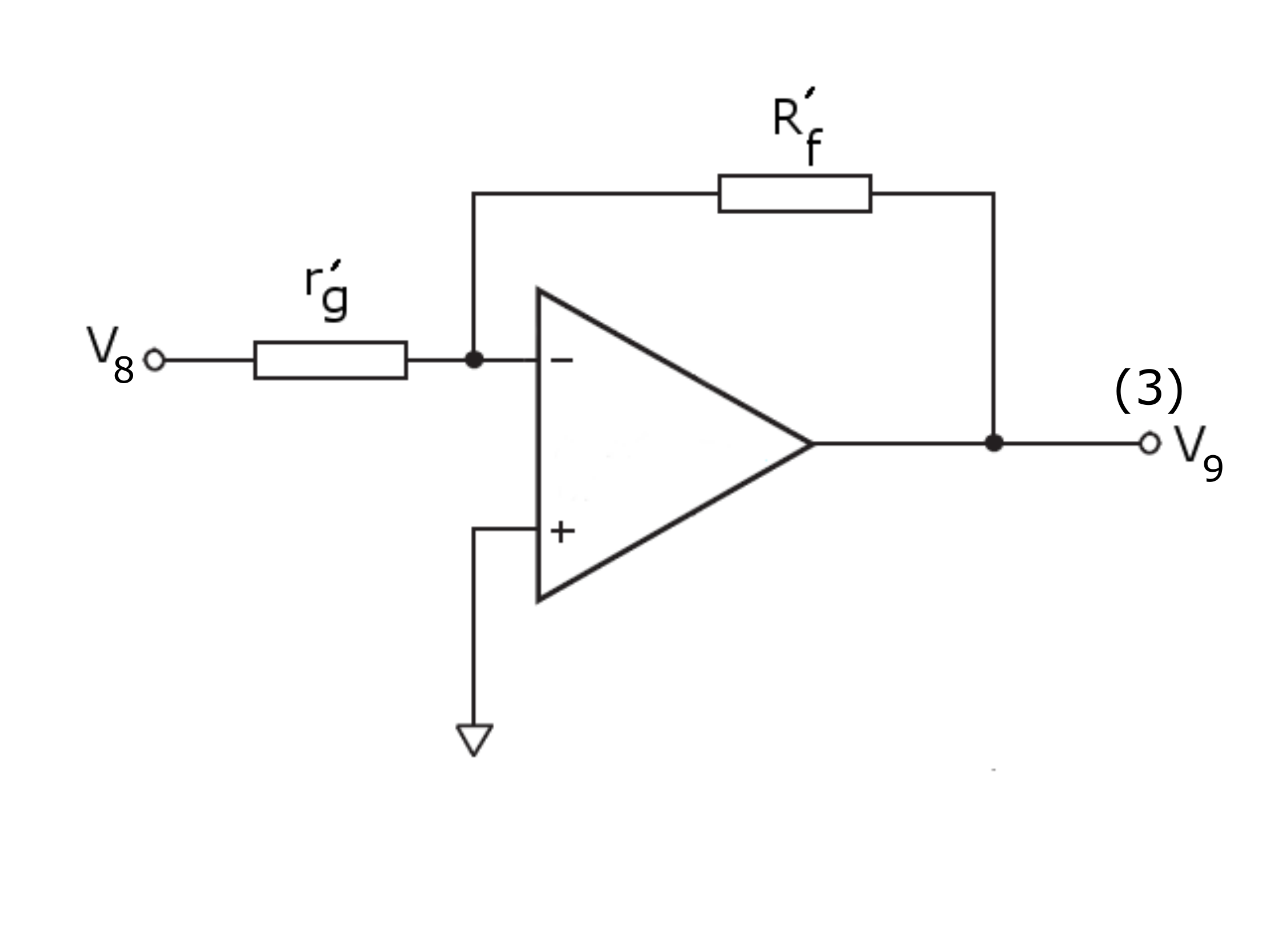}
\caption{Инвертиращ усилвател}
\label{bInverting amplifier}
\end{minipage}
\begin{minipage}[c]{0.57\linewidth}
\includegraphics[scale=0.28]{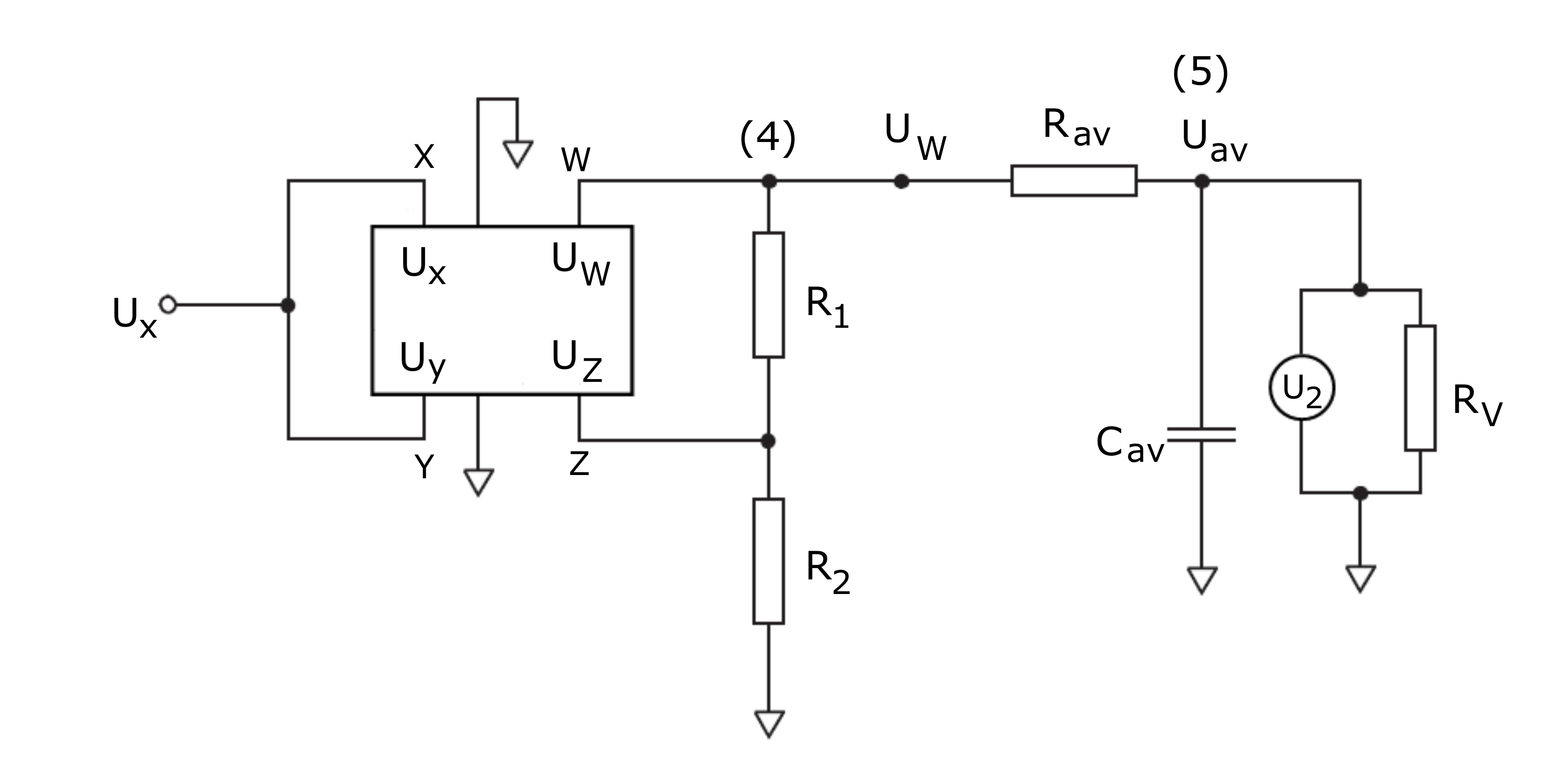}
\caption{Умножител на напрежения}
\label{bAnalog multiplier}
\end{minipage}
\end{figure}

\section{Задача за домашно, което трябва да се изпрати в нощта след олимпиадата до изгрев слънце на електронния адрес на олимпиадата: 
Изведете използваните формули за определяне на константата на Болцман $k_\mathrm{_B}$. Зомерфелдова премия с паричен еквивалент от DM137. XL}

Упътване. Започнете с теоремата за равноразпределението, приложена към енергията на топлинните флуктуации 
$\frac12 C\left< U^2 \right>=\frac12 k_\mathrm{_B}T.$
Като анализирате електронната схема на постановката,
изразете средно квадратичното напрежение $\left<U^2\right>=U_2 U^* $ чрез експериментално измерваното напрежение $U_2$.
Трябва да се изрази коефициента $U^*$ във формулата за линейната регресия
$U_2 U^* /T= \left<U^2\right>/T = k_\mathrm{_B}\, (1/C) + \mathrm{const},$
от която по наклона на правата се определя константата на Болцман $k_\mathrm{_B}$
Накратко, напишете статия как идеята на Айнщайн за определяне на константата на Болцман се реализира с нашата постановка.
Посочете какви подобрения на постановката биха повишили точността на измерването.

Можете да работите в колектив от участници в Олимпиадата, 
може да се консултирате с професионалисти,
но трябва да се посочат имената на консултантите и пълния списък на авторите.

Най-добрата работа се награждава със Зомерфелдова премия с паричен еквивалент от DM137. 
Наградата се дава само лично и само в деня на обявяването на резултатите
10.12.2017.

\section{Задачи за по-нататъшна работа}

Олимпиадата е замислена да помогне и на ученици, лишени от качествено образование. 
Ученици идващи от училища, където кабинетите по физика са затворени. 
Затова не се смущавайте, ако не сте доволни от постигнатите резултати.
Попитайте учителя си дали подходящ уред за да измерите кондензаторите по-точно.
Спокойно повторете експеримента вкъщи и ще видите колко е било просто.
Щом можете вкъщи да измерите фундаментална константа, значи 
имате летящ старт за всички природни и технически науки.
Ако имате възможност, например в кръжок по физика, може да демонстрирате 
как работи постановката в училище.
Моля, напишете ни отзив с вашето мнение за Олимпиадата.
Ние организаторите ще оценим високо вашите препоръки,  
които ще се опитаме да отчетем за следващото издание на олимпиадата
EPO6, догодина по същото време.

Постановката от Олимпиадата е наистина уникална.
Разполагате с усилвател, който усилва милион пъти, с ниво на шума от 
$1\;\mathrm{nV/\sqrt{Hz}}.$

С подходяща модификация постановката може да се използва и за 
измерване на заряда на електрона и абсолютната температура.
Нещо повече, с минимално изменение постановката от Олимпиадата е 
всъщност един локин (lock-in) волтметър, 
който може да се използва за измерване на променливи напрежения 
даже по-малки от 1 микроволт ($\mathrm{1\;\mu V=10^{-6}\;V}$).
Именно заради тази възможност от постановката излиза жичката ``NC'', 
която не се използва в описаните задачи.
За детайли следете публикациите на авторите на сървъра, 
където са публикувани задачите и решенията от предишните Олимпиади.

\appendix

\section{Снимка на експерименталната постановка}

\begin{figure}[h]
\centering
\includegraphics[scale=0.25]{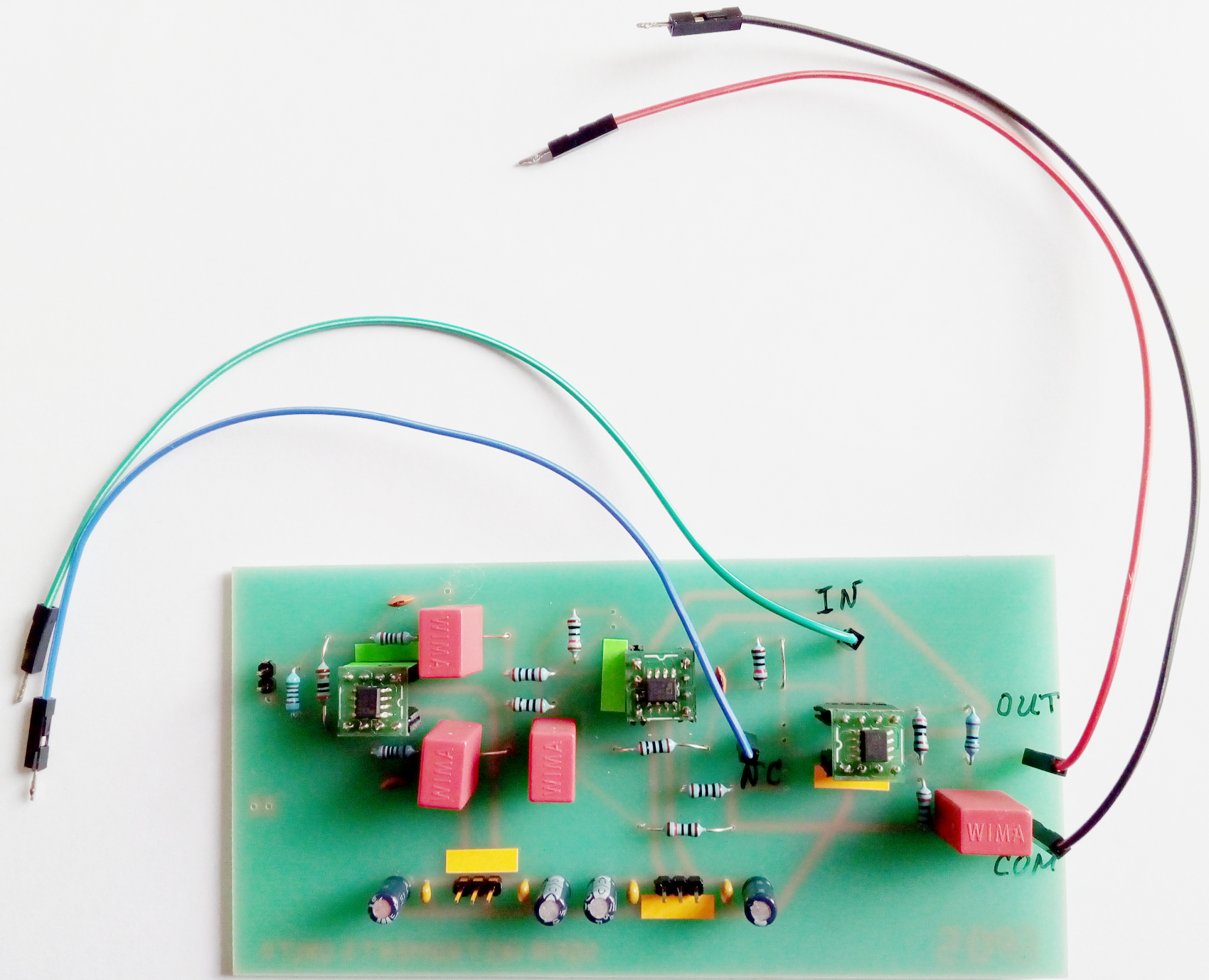}
\caption{Снимка на постановката. В дясно се виждат изводите и жиците между които се измерва
разликата $U_2$ между изходното напрежение ``OUT'' и земята ``COM''. 
Входното напрежение $U_1$ се подава между входа (IN) и земята ``COM''.
Изводът ``NC'' може да се използва за по-нататъшни изследвания.
На рейката най-отляво се поставят различните кондензатори $C_n$.
}
\label{Fig:bPhoto}
\end{figure}

\section{Решения на експерименталните задачи}

\subsection{Начални лесни задачи. S}
\begin{enumerate}
\item $U_{\mathrm{B}1}=9.72$~V, $U_{\mathrm{B}2}=9.73$~V, $U_{\mathrm{B}3}=9.73$~V, $U_{\mathrm{B}4}=9.73$~V, а за измерването на батерията от 1.5~V трябва да се ползва обхвата на волтметъра, който дава най-добра точност (обикновено този обхват е 2~V),
$U_{\mathrm{B}5}=1.582$~V.
\item Потенциометърът има три контакта, като крайните два са запоени за държача на батерията.
Потенциометрично свързване означава свързване на средния и един от крайните контакти,
$U_\mathrm{min}=-1.577$~V, $U_\mathrm{max}=1.577$~V и следователно интервалът от напрежения е $U \epsilon [-1.577,1.577]$~V.
\item Просто поставяне на 9~V батерии в техните държачи.

\subsection{Аналогово повдигане на квадрат. M}
\item Ориентиране на платката според инструкциите.
\item Включване на напрежение според инструкциите.
\item Свързване на батерията към съответните изводи на платката според инструкциите.
\item Успоредно свързване на първия волтметър на входа според инструкциите.
\item Проверка за правилно свързване.
\item Свързване на втория волтметър на изхода на постановката според инструкциите.
\item Проверка за правилно свързване на двата волтметъра, чиито ``COM'' електроди трябва да са свързани.
\item Таблица~\ref{tbl:bmult} без последната колона (задача 13).
\begin{center}
\begin{table}[h]
\begin{tabular}{ c  r  r  r }
		\tableline \tableline
		&  \\ [-1em]
		i & $U_1$ [V] & \hspace{2.5pt} $U_2$ [V] & \hspace{5pt} $U_1^2$ [V$^2$] \\ \tableline 
			&  \\ [-1em] 
			1	&	0.1	&	0.011	&	0.01	\\
			2	&	0.2	&	0.023	&	0.04	\\
			3	&	0.4	&	0.07	&	0.16	\\
			4	&	0.6	&	0.15	&	0.36	\\
			5	&	0.8	&	0.263	&	0.64	\\
			6	&	1.0	&	0.386	&	1.00	\\
			7	&	1.2	&	0.551	&	1.44	\\
			8	&	1.4	&	0.748	&	1.96	\\
			9	&	1.573  &	0.941	&	2.474	\\
			10	&	-1.572	&	0.942	&	2.471	\\
			11	&	-1.4	&	0.75	&	1.96	\\
			12	&	-1.2	&	0.553	&	1.44	\\
			13	&	-1.0	&	0.387	&	1.00	\\
			14	&	-0.8	&	0.265	&	0.64	\\
			15	&	-0.6	&	0.151	&	0.36	\\			
			16	&	-0.4	&	0.071	&   0.16	\\
			17	&	-0.2	&	0.024	&	0.04	\\
			18	&	-0.1	&	0.012	&	0.01	\\			
\tableline \tableline
\end{tabular}
	\caption{Резултати от измерванията в задача 11 и обработката на резултатите в задача 13 (последната колона).}
	\label{tbl:bmult}
\end{table}
\end{center}
\item Резултатите са представени на Фиг.~\ref{fig:bpar}.
\begin{figure}[h]
\begin{minipage}[t]{0.48\linewidth}
\centering
\includegraphics[scale=0.64]{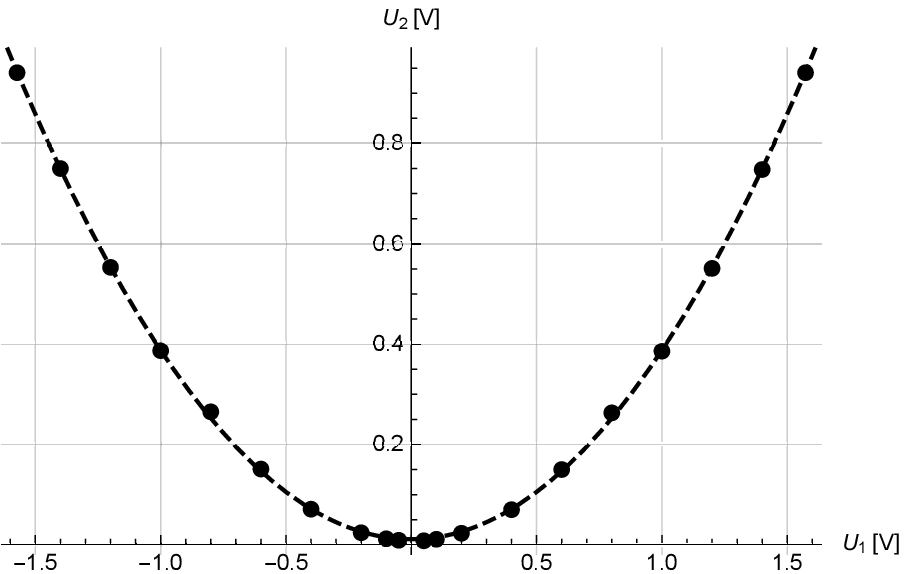}
\caption{Графично представената зависимост $U_2$ от  $U_1$ от Таблица~\ref{tbl:bmult}.}
\label{fig:bpar}
\end{minipage}
\hfill
\begin{minipage}[t]{0.48\linewidth}
\centering
\includegraphics[scale=0.64]{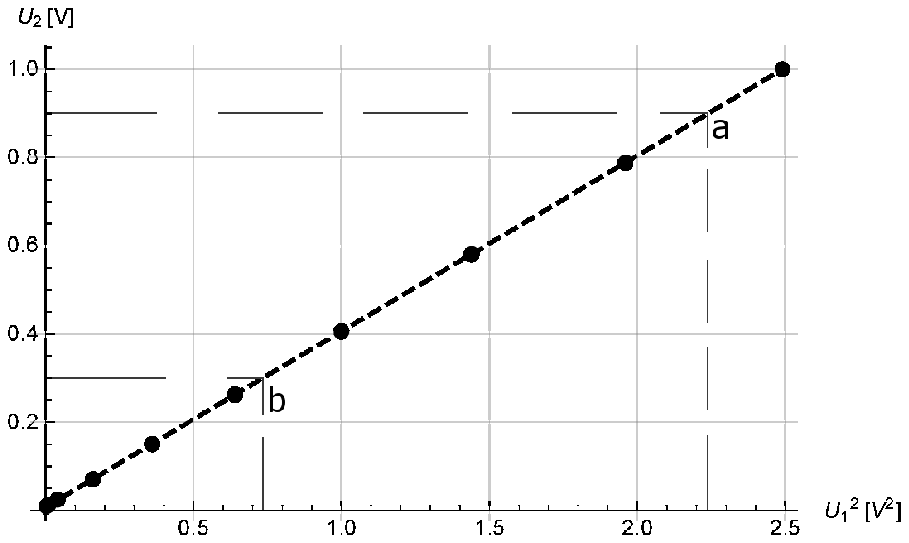}
\caption{Графично представената зависимост $U_2$ от $(U_1)^2$ от Таблица~\ref{tbl:bmult}.}
\label{fig:blin}
\end{minipage}
\end{figure}
\item Последната колона на Таблица~\ref{tbl:bmult} и резултатите са представени на Фиг.~\ref{fig:blin}.
Избираме 2 точки (a) и (b) на Фиг.~\ref{fig:blin}, така че по ордината $(U_2)_a=0.9$~V и $(U_2)_b=0.3$~V и намираме съответните им стойности по абсциса $(U_1)_a^2 \approx 2.24$~V$^2$ и $(U_1)_b^2 \approx 0.725$~V$^2$, като за всяка от двете точки чертаем права успоредна на ординатата към абсцисата, и пресечната им точка е съответната стойност за $(U_1)^2$
\be
U_0=\frac{\Delta(U_1^2)}{\Delta(U_2)}=
\frac{(U_1)_a^2-(U_1)_b^2}{(U_2)_a-(U_2)_b}=\frac{2.24-0.725}{0.9-0.3}=2.525~\mathrm{V}.
\ee

\subsection{Флуктоскопия на напрежението. Определяне на константата на Болцман. L}

\item Разкачване на входните източник на напрежение и волтметър според инструкциите. 

\item Подаване на напрежение към усилвателя \textbf{според инструкциите}.

\item Свързване на двата операционни усилвателя към платката \textbf{според инструкциите}.

\item Стойностите на капацитетите на набора от кондензатори е нанесен в първите 2 колони на Таблица~\ref{tbl:bmeas}.
\begin{center}
\begin{table}[h]
\begin{tabular}{ c  r  r  r  r }
		\tableline \tableline
		&  \\ [-1em]
		$n$  \hspace{1.5pt} & $C_n$ [nF] & \hspace{5pt} $U_2$  [V] & \hspace{1.5pt} $x_n$ [1/$\mu$F] & \hspace{1.5pt} $y_n$ [f\,V$^2$/K]\\ \tableline 
			&  \\ [-1em] 
			1	&	14.8	&	0.293	&	67.57 	&	2.458 \\
			2	&	27.9	&	0.255	&	35.84	&	2.139 \\
			3	&	33.4	&	0.244	&	29.94	&	2.047 \\
			4	&	42.3	&	0.233	&	23.64	&	1.955 \\
			5	&	66.9	&	0.202	&	14.95	&	1.695 \\
			6	&	105.0	&	0.204	&	9.524	&	1.712 	\\
\tableline \tableline
\end{tabular}
	\caption{Експериментални резултати от измерването на напрежението на топлинните 				флуктуации за набора кондензатори. 
	Тук за краткост сме въвели префикса $\text{f}=10^{-15}$.}
	\label{tbl:bmeas}
\end{table}
\end{center}
\item Резултатите от измерването според инструкциите са нанесени в 3-тата колона на Таблица~\ref{tbl:bmeas}.

\item За $U_0$ получихме 2.525~V, следователно
\be
U^*=\frac{U_0}{Y^2}=\frac{2.525}{(1.01 \times 10^6)^2}=\frac{2.525}{(1.01)^2 \times 10^{12}}=
2.475 \times 10^{-12}=2.475~\mathrm{pV}.
\ee

\item За преценка (не точно измерване) на температурата в аудиторията не е необходим термометър.
За стайна температура може да се приемат стойности между 20$^\circ$C и 25$^\circ$C, които
в Келвини (добавяме още 273$^\circ$C) са съответно 293~K и 298~K, ние ще изберем по-малката стойност $T=293$~K.
Разликата между двете стойности е под 2\%, което е напълно задоволително за нашата задача.
Още повече, тези 5$^\circ$ са напълно осезаеми и на практика всеки човек е способен да различи още по-малки температурни разлики (например дали е топло в аудиторията, за да работите по тениска).
Вашето тяло е един термометър и както всеки измерителен прибор, за надеждни показания е необходима правилна калибровка.

\item Резултатите се намират в последните 2 колони на Таблица~\ref{tbl:bmeas}, като вторият индекс $n$ на означението $U_{2,n}$ е номерът на съответния кондензатор, а представката ``f'' на  fV$=10^{-3}~\mathrm{pV}=10^{-15}$~V се нарича фемто-волт.

\item Преценете мащаба за графичното представяне на получените измервания. 

\item Правата е начертана на Фиг.~\ref{fig:buc}.

\begin{figure}[h]
\centering
\includegraphics[scale=1.0]{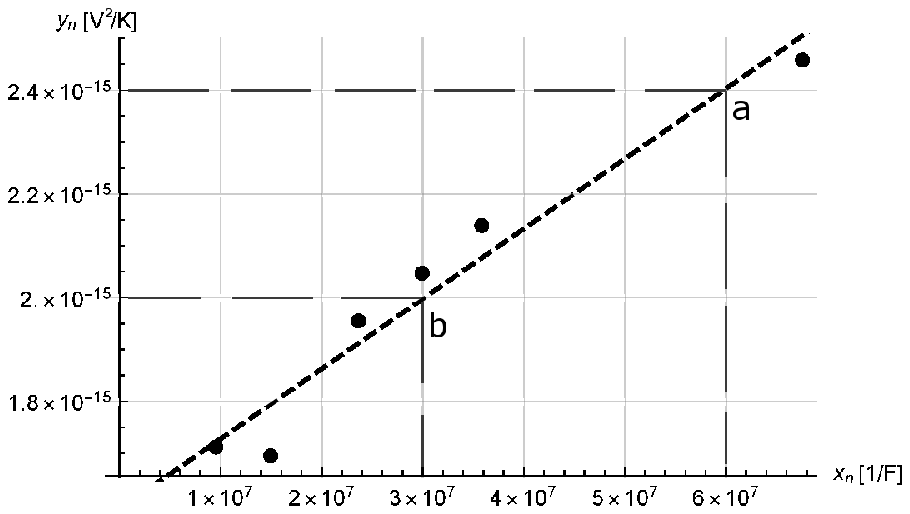}
\caption{Графично представената зависимост $y_n$ от $x_n$ от Таблица~\ref{tbl:bmeas}.
Първоначално са нанесени експерименталните точки.
После се прекарва приближаващата права.
Тя може да се начертае и на око, 
а математичната процедура се нарича линейна регресия.
От правата избираме 2 точки (a) и (b) и пресмятаме наклона
$k_\mathrm{_B}=\Delta y/ \Delta x$, който в нашия случай дава константата на Болцман.}
\label{fig:buc}
\end{figure}

\item Аналогично на задача 13, си избираме 2 точки (a) и (b) от правата  с координати съответно $x_a= 6 \times 10^7~$F$^{-1}$, $y_a=2.4 \times 10^{-15}$~V$^2$/K и $x_b= 3 \times 10^7~$F$^{-1}$, $y_b=2 \times 10^{-15}$~V$^2$/K и пресмятаме разликите в координатите
\begin{align}
& \Delta y=y_a-y_b=(2.4-2) \times 10^{-15} = 0.4 \times 10^{-15} \, \mathrm{V^2/K}, \\ 
& \Delta x=x_a-x_b=(6-3) \times 10^7 = 3  \times 10^7 \, \mathrm{1/F}.
\end{align}

\item С получените стойности за разликите в координатите, пресмятаме наклона, който всъщност е $k_\mathrm{_B}$
\be
k_\mathrm{_B}=\frac{\Delta y}{\Delta x}=
\frac{0.4}{3} \times 10^{-22}=1.33 \times 10^{-23} \,( \mathrm{F\,V^2/K}=\mathrm{J/K}).
\ee

Нека отчетем, че константата на Болцман е много малка и даже правилното определяне на нейния порядък с толкова проста постановка е значителен успех. 
Когато сигналът се усилва милион пъти по напрежение 
и $10^{12}$ пъти (120~dB) по мощност
един множител от 2 (3~dB) означава 
грешка в логаритмичен мащаб 3/120=2.5~\%,
което е значителен успех за постановка предназначена за средното образование.
Все пак се случва постановката дава точност от няколко процента.
При добре известната експериментална стойност
$k_\mathrm{_B}=1.38064 \times 10^{-23}$~J/K
относителната грешка на измерването е
$(138-133)/138<4\%.$
Систематична грешка дават ефектите на неидеалност на операционните усилватели,
но основен източник на неопределеност е плаването на нулата на измервателната постановка.
Намаляването на дрейфа на нулата на постановката е възможно, 
но значително би усложнило схемата до ниво на университетска физика или даже метрология.

\end{enumerate}

\section{Няколко прости задачи за закона на Ом}

Нека разгледаме 6 последователни задачи, които се решават чрез прилагане на закона на Ом.
На Фигури~\ref{bNon-inverting amplifier}-\ref{bInverting amplifier} токът, който излиза от десния ъгъл на триъгълника е такъв, че напреженията в точките (+) и (-) са равни
\begin{equation}
V_+-V_-=0,
\label{beqpoten}
\end{equation}
а в тези входове токът е пренебрежим.
\begin{enumerate}
\item Нека разгледаме тока $I_2=V_2/(R_\mathrm{f}+r_\mathrm{g})$, който се спуска от изхода $V_2$ до ``земята'' 
(\begin{circuitikz}
\draw (0,0) node[sground,xscale=0.25,yscale=0.45]{};
\end{circuitikz})
с потенциал 0.
По закона на Ом напрежението в точката между резисторите $R_\mathrm{f}$ и $r_\mathrm{g}$, което е същото както и на (-) входа е
\begin{equation}
V_-=r_\mathrm{g} I_2 = \frac{r_\mathrm{g}}{R_\mathrm{f}+r_\mathrm{g}}V_2.
\end{equation}
От друга страна, за всеки от триъгълниците двете напрежения са почти равни и
\begin{equation}
V_1=V_+=V_-=\frac{r_\mathrm{g}}{R_\mathrm{f}+r_\mathrm{g}}V_2.
\end{equation}
Така получаваме за коефициента на усилване
\begin{equation}
y_1=\frac{V_2}{V_1}=\frac{R_\mathrm{f}}{r_\mathrm{g}}+1.
\end{equation}
Тъй като $y_1>0$ този усилвател се нарича неинвертиращ.
Електронните устройства, които изравняват потенциалите на двата входа съгласно Ур.~(\ref{beqpoten}) се наричат операционни усилватели, но терминологията само отблъсква начинаещите.
\item На Фиг.~\ref{bBuffer} са начертани два такива усилвателя, за които
\begin{align}
& V_2=y_1 V_1, \nn \\
& V_4=y_1 V_2.
\end{align}
Ако извадим едното уравнение от другото, получаваме същия коефициент на усилване
\begin{equation}
\frac{V_2-V_4}{V_1-V_2}=y_1=\frac{R_\mathrm{f}}{r_\mathrm{g}}+1.
\end{equation}
Земята на Фиг.~\ref{bBuffer} е нарисуване неправилно, защото тази точка не е необходимо да има нулев потенциал и се оставя свободна.
Такава земя се нарича виртуална и вместо две последователно свързани съпротивления $r_\mathrm{g}$ се използва едно двойно $r_\mathrm{_G}=2 r_\mathrm{g}$ и тогава
\begin{equation}
y_1=1+\frac{2 R_\mathrm{f}}{r_\mathrm{_G}}.
\end{equation}
Такава схема с два неинвертиращи усилватели се нарича буфер.
На буфера, начертан на Фиг.~\ref{fig:bcircuit}, параметрите на схемата са дадени в Таблица~\ref{tbl:bvalues}, където
$r_\mathrm{_G} =20~\Omega, R_\mathrm{f}\equiv R_\mathrm{F}=1~\mathrm{k}\Omega$ и $y_1 \approx 101$.
Използваме резистори с 1\% точност.
\item Усилвателят показан на Фиг.~\ref{bDifferential amplifier} се анализира също като се разглеждат делители на напрежение, описани от закона на Ом.

От точка с потенциал $V_6$ до земята се спуска ток
$I_6=V_6/(R^\prime_\mathrm{f}+r^\prime_\mathrm{g})$.
По закона на Ом потенциалът в точка (+) е
\be
V_+=0+R^\prime_\mathrm{f} I_6 = \frac{R^\prime_\mathrm{f}}{R^\prime_\mathrm{f}+r^\prime_\mathrm{g}} V_6.
\label{eq:bd-}
\ee
Съвършено аналогично от точка $V_5$ към $V_7$ се спуска ток $I_5=(V_5-V_7)/(R^\prime_\mathrm{f}+r^\prime_\mathrm{g})$ и потенциалът в точка (-) e
\be
V_-=V_7+\frac{V_5-V_7}{R^\prime_\mathrm{f}+r^\prime_\mathrm{g}}R^\prime_\mathrm{f}.
\label{eq:bd+}
\ee
Токът от триъгълника (символизиращ операционния усилвател) е такъв, че $V_+=V_-$ и заместването от двете по-горни уравнения (\ref{eq:bd-} и \ref{eq:bd+}) дава
\be
\frac{R^\prime_\mathrm{f}}{R^\prime_\mathrm{f}+r^\prime_\mathrm{g}} V_6=
V_7\frac{R^\prime_\mathrm{f}+r^\prime_\mathrm{g}}{R^\prime_\mathrm{f}+r^\prime_\mathrm{g}} +\frac{V_5-V_7}{R^\prime_\mathrm{f}+r^\prime_\mathrm{g}}R^\prime_\mathrm{f}.
\label{eq:bd=}
\ee
В дясната страна на това Ур.~(\ref{eq:bd=}) членовете в числителите, които имат
$R^\prime_\mathrm{f} V_7$ се съкращават.
Умножаваме горното уравнение с $(R^\prime_\mathrm{f}+r^\prime_\mathrm{g})$ и получаваме
\be
R^\prime_\mathrm{f} V_6=r^\prime_\mathrm{g} V_7 + R^\prime_\mathrm{f} V_5,
\ee
което може да се препише като
\be
y_2=\frac{V_7}{V_6-V_5}=\frac{R^\prime_\mathrm{f}}{r^\prime_\mathrm{g}}.
\label{eq:bdelta}
\ee
Съгласно параметрите от Таблица~(\ref{tbl:bvalues})
$R^\prime_\mathrm{f} \equiv R^\prime_\mathrm{F} = 10~\mathrm{k}\Omega$,
$r^\prime_\mathrm{g} \equiv R_\mathrm{G} = 100~\Omega$ и $y_2 \approx 100$.
Формула~(\ref{eq:bdelta}) описва усилването на един разликов усилвател.
Буфер, последван от разликов усилвател се нарича инструментален усилвател.
За разгледания от нас инструментален усилвател $y_1 y_2 \approx 10100$ и така постепенно навлизаме в терминологията, която се използва в електрониката.
\item На Фиг.~\ref{bInverting amplifier} е представен един усилвател, който може да се разглежда като разликов усилвател със заземен (+) вход.
Тогава от Ур.~(\ref{eq:bdelta}) получаваме
\be
y_3=\frac{V_9}{V_8}=-\frac{R^\prime_\mathrm{f}}{r^\prime_\mathrm{g}}.
\ee
Заместването на параметри от Таблица~(\ref{tbl:bvalues})
$R^\prime_\mathrm{f} \equiv R^\prime_\mathrm{F} = 10~\mathrm{k}\Omega$,
$r^\prime_\mathrm{g} \equiv R_\mathrm{G} = 100~\Omega$ дава $y_3 = 100$
и така пълното усилване на нашата схема е
\be
Y=y_1 y_2 y_3 \approx -1.01 \times 10^6.
\ee

Досега разглеждахме случаи на статика, за които имаме прави напрежения и токове, или с други думи при нулева честота $f=0$.
На Фиг.~\ref{fig:circuit}, обаче, имаме кондензатори $C_\mathrm{G}$ последователно свързани на резисторите $R_\mathrm{G}$,
и кондензатори $C_\mathrm{F}$ и $C^\prime_\mathrm{F}$ успоредно свързани съответно на резисторите $R_\mathrm{F}$ и $R^\prime_\mathrm{F}$.
Въпреки това, това усилване от около милион пъти е приложимо в честотния интервал, за който
\be
\frac{1}{\omega C_\mathrm{G}} \ll R_\mathrm{G},
\ee
и същевременно с това
\be
\omega C^\prime_\mathrm{F} \ll \frac{1}{R^\prime_\mathrm{F}},
\ee
където $\omega= 2 \pi f$ е кръгова честота.
Неравенствата описват пренебрежимо влияние на кондензаторите.
Големият кондензатор $C_\mathrm{G}$ трябва да има пренебрежим импеданс в сравнение със съпротивлението $R_\mathrm{G}$.
И едновременно с това, малкият кондензатор $C^\prime_\mathrm{F}$ трябва да има пренебрежима проводимост в сравнение с проводимостта на съпротивлението $R^\prime_\mathrm{F}$.
И накрая, при фиксирано съпротивление $R$ на входа на усилвателя за целия набор от различни входни кондензатори $C_\mathrm{i}$, съответната времеконстанта $RC_\mathrm{i}$ трябва да попада в оптималния интервал, в който работи усилвателя
\be
R^\prime_\mathrm{F} C^\prime_\mathrm{F} \ll RC_\mathrm{i} \ll R_\mathrm{G} C_\mathrm{G}.
\ee
Тъй като
$R_\mathrm{G} C_\mathrm{G}/R^\prime_\mathrm{F} C^\prime_\mathrm{F} \approx 10^4 \gg 1$,
изискваните условия са удовлетворени и корекциите към опростеното разглеждане при нулева честота са само няколко процента.
\item Нека разгледаме и нелинейния елемент, който осигурява измерването на средноквадратичното напрежение на усиления сигнал $U_X$.
На Фиг.~\ref{bAnalog multiplier} е показано устройство, за което напрежението на изхода $U_W$ и напреженията на входовете $U_X, U_Y$ и $U_Z$ са свързани с релацията
\be
U_W=\frac{U_X U_Y}{U_\mathrm{m}}+U_Z,
\label{eq:bmult}
\ee
където за използвания в постановката умножител AD633 константата $U_\mathrm{m}=10$~V.
Токът $I_W = U_W/(R_1+R_2)$, който се спуска от точката W към земята създава в точката Z напрежение
\be
U_Z=I_W R_2 = \frac{R_2}{R_1+R_2} U_W.
\ee
Заместваме $U_Z$ в Ур.~(\ref{eq:bmult}) и допълнително отчитайки, че съгласно свързването от Фиг.~\ref{bAnalog multiplier} $U_Y=U_X$, получаваме
\begin{align}
& U_W=\frac{U_X^2}{U_\mathrm{m}}+\frac{R_2}{R_1+R_2} U_W \quad \mbox{или} \nn \\
& U_W=\frac{U_X^2}{U_\mathrm{m}}\frac{R_1+R_2}{R_1}= U_X^2/\tilde U, \quad
\tilde U = U_\mathrm{m} \frac{R_1}{R_1+R_2}.
\label{eq:buw}
\end{align}
\item Накрая схемата завършва с усредняващ филтър с голяма времеконстанта
$R_\mathrm{av}C_\mathrm{av} = 15$~сек. 
След този филтър наблюдаваме усредненото по време напрежение $\left<U_W\right>$.
Това средно напрежение през усредняващия резистор $R_\mathrm{av}$ и волтметъра с вътрешно съпротивление $R_\mathrm{V}$ създава среден ток
\be
I_\mathrm{V}=\frac{\left < U_W \right >}{R_\mathrm{av}+R_\mathrm{V}}
\ee
и средното напрежение, което измерва волтметъра е
\be
U_\mathrm{V}=\frac{R_\mathrm{V}}{R_\mathrm{av}+R_\mathrm{V}} \left < U_W \right >.
\label{eq:buv}
\ee
Така, комбинирайки горното уравнение с Ур.~(\ref{eq:buw}) получаваме
\be
U_\mathrm{V}=\frac{\left < U_X(t)^2 \right >}{U_\mathrm{m}} \frac{R_1+R_2}{R_1}
\frac{R_\mathrm{V}}{R_\mathrm{av}+R_\mathrm{V}} =
\frac{\left < U_X(t)^2\right >}{U_0},
\label{eq:bmeas}
\ee
където за краткост въвеждаме означението
\be
U_0 = \frac{R_1}{R_1+R_2}\frac{R_\mathrm{av}+R_\mathrm{V}}{R_\mathrm{V}}U_\mathrm{m}.
\ee
Параметърът с размерност напрежение $U_0$ може да се измери експериментално, ако в точката X подаваме с потенциометър различни постоянни напрежения и измерваме напрежението на волтметъра.

\end{enumerate}

Сега вече може да синтезираме работата на цялата постановка.
Топлинното напрежение на кондензатора на входа $U(t)$ се усилва
\be
Y=y_1 y_2 y_3 = - \left ( \frac{2 R_\mathrm{F}}{r_\mathrm{_G}} + 1\right )
\frac{R^\prime_\mathrm{F}}{R_\mathrm{G}} \frac{R^\prime_\mathrm{F}}{R_\mathrm{G}}
\label{eq:bfinal}
\ee
пъти $U_X(t)=YU(t)$, след което се повдига на квадрат, усреднява и измерва с волтметър
\be
U_\mathrm{V}=\frac{\left < U_X(t)^2 \right >Y^2}{U_0}=
\frac{\left <U_X(t)^2\right >}{U^*},
\ee
където
\be
U^* \equiv \frac{U_\mathrm{m}}{Y^2}  \frac{R_1}{R_1+R_2}\frac{R_\mathrm{av}+R_\mathrm{V}}{R_\mathrm{V}}=\frac{U_0}{Y^2}.
\ee
Така, ако теоремата за равноразпределението дава
\be
C \left < U(t)^2 \right > = k_\mathrm{_B} T,
\ee
заместването в Ур.~(\ref{eq:bfinal}) дава
\be
U_\mathrm{V}=\frac{k_\mathrm{_B} T}{CU^*} + \mathrm{const}.
\ee
Така получената формула се използва за измерването на константата на Болцман в подзадачи~23-25, където $y=U_\mathrm{V}$ и $x=T/CU^*$.

В заключение, комбинирайки 6 прости задачи от закона на Ом, ние получихме пълната теория на работата на експерименталната постановка и методът за обработването на получените експериментални данни.

\section{ОЕФ5}

Тази задача бе дадена на 5-тата Олимпиада по експериментална физика и повече от 137 от участвалите ученици знаеха закона на Ом и можаха не само да разберат след олимпиадата, но и по време на състезанието да извършат елементарните пресмятания.
Освен закона на Ом като математика, ние използвахме само решаването на линейно уравнение и елементарни замествания.
И съвсем не е нужно да си специалист с практика в областта на аналоговата електроника, за да анализираш схеми, на които в точките (+) и (-) на някакъв триъгълник, потенциалите са еднакви.
Учениците с лекота решават серия от 6 задачи от закона на Ом
и съвсем не се нуждаят от курс по аналогова електроника.
Нещо повече, учителите слушали някакъв подобен курс не могат да си спомнят нещо полезно за
преподаването и специално за решаването на задачата.
В Олимпиадата участваха вдъхновени ученици и от Охрид и от Астана, а ако трябва да се сравняваме с други континенти, разстоянието между тези два града е по-голямо
от разстоянието между Лос Анджелес и Ню Йорк.

\section{Напредничава електроника приложена към нашата постановка}

И все пак на инженерно или поне университетско ниво, ако желаем да постигнем 1\% точност, се налага да отчетем честотната  зависимост, възникваща от основното уравнение на операционните усилватели в честотно представяне
\be
U_+-U_-=\varepsilon U_0, \quad \varepsilon(\omega) \approx \mathrm{j} \omega \tau, \quad
\mathrm{j}^2=-1, \quad \tau = \frac{1}{2 \pi f_0},
\label{eq:bmaster}
\ee
където $f_0$ е срязващата честота на операционния усилвател.
Определянето на $f_0^\text{(ADA4898-2)} \approx 40$~MHz е описано в литературен 
извор~\onlinecite{Manhattan}.
Подробно пресмятане на честотните зависимости на 
коефициентите на усилване на усилвателите е дадено в друг извор ~\cite{Detailed}, тук просто ще напишем само техните стойности
\begin{align}
& \Upsilon_1=\frac1{Y^{-1}(\omega)+\varepsilon(\omega)},
\quad Y(\omega)=\frac{Z_\mathrm{f}(\omega)}{z_\mathrm{g}(\omega)}+1, \quad
\Upsilon_1(0)=y_1=\frac{R_\mathrm{f}}{r_\mathrm{g}}+1, \quad
z_\mathrm{g}=r,
\quad \frac1{Z_\mathrm{f}}=\frac1{R_\mathrm{f}}+\mathrm{j}\omega C_\mathrm{f}, \\
& \Upsilon_2= \frac1{\Lambda(\omega)+\varepsilon(\omega)\left[\Lambda(\omega) +1\right]}, \quad
\Lambda(\omega)=\frac{z_\mathrm{g}(\omega)}{Z_\mathrm{f}^\prime(\omega)}, \quad
\Upsilon_2(0)=y_2=\frac{R_\mathrm{f}^\prime}{r_\mathrm{g}^\prime},
\quad z_\mathrm{g}=r_\mathrm{g}^\prime+\frac1{\mathrm{j}\omega C_\mathrm{g}},
\quad Z_\mathrm{f}^\prime=R_\mathrm{f}^\prime, \\
& \Upsilon_3=-\frac{1}{\mathcal{L}(\omega)+\varepsilon(\omega)\left[\mathcal{L}(\omega)+1\right]},
\quad \mathcal{L}(\omega)=\frac{z_\mathrm{g}(\omega)}{Z_\mathrm{f}^{\prime\prime}(\omega)},
\quad \Upsilon_3(0)=y_3=-\frac{R_\mathrm{f}^\prime}{r_\mathrm{g}^\prime},
\quad \frac{1}{Z_\mathrm{f}^{\prime\prime}}=\frac{1}{R_\mathrm{f}^\prime}+
\mathrm{j} \omega C_\mathrm{f}^\prime,
\end{align}
и за целия честотно-зависим коефициент на усилване на усилвателя получаваме
произведението на отделните независими усилватели
\be
\Upsilon(\omega)=\Upsilon_1 \Upsilon_2 \Upsilon_3.
\ee
След пресмятане на ширината на лентата на усилвателя, 
което е важно понятие в електрониката извеждаме правилото на заместване
\be
\frac{1}{U^*}=\frac{1}{U_0}Y^2 \rightarrow \frac{1}{U_0} Y^2 Z,
\ee
където корекционният множител Z се дава с интеграла
който описва ширината на лентата на усилвателя и филтъра поставен на входа му
\be
Z=1+\epsilon
=\dfrac{\int_0^\infty \frac{\left |\Upsilon(\omega)\right |^2}{1+(\omega RC)^2} \mathrm{d}\omega}
{ \int_0^\infty \frac{Y^2}{1+(\omega RC)^2} \mathrm{d}\omega}
=\frac{2}{\pi} \dfrac{RC}{Y^2} \int_0^\infty \frac{\left |\Upsilon(\omega)\right |^2}{1+(\omega RC)^2} \mathrm{d}\omega.
\ee
Интеграла от знаменателя дава ширината на лентата, ако усилвателя работи до много високи честоти и не са поставени кондензаторите които спират плаването на нулите и самовъзбуждането. 
Численото пресмятане за корекцията дава $\epsilon \approx 6\%$, с което се отчитат всички систематични грешки при използването на нашата постановка.
В тази формула трябва да бъдат замествани капацитетите $C_n$ от изследваната серия.

\section{Локин волтметър скрит в експерименталната постановка}

Нека сега обърнем внимание на елемента с наименование S\textunderscore M във Фиг.~\ref{bPCB} и \ref{fig:bcircuit}.
В схемата на Фиг.~\ref{fig:bcircuit} S\textunderscore M е ключ, чието първоначално положение е затворено, което изравнява потенциалите на двата входа ``IN'' и ``NC''.
Или с други думи, $U_X=U_Y$, както вече разгледахме при работата на нелинейния елемент.

За задачите на Олимпиадата ключът S\textunderscore M не е необходим и затова не е поставен на платката на Фиг.~\ref{bPCB}.
Вместо това, двата контакта на ключа са запоени, което на практика означава, че ключът е постоянно затворен или окъсен (т.е. липсващ).
Спойката свързваща двата контакта на ключа може лесно да се премахне като се отреже със секачки или се разпои, като по този начин ключът се отваря и вече $U_X \neq U_Y$, което означава че умножителят умножава два различни сигнала.

Нека сега разгледаме принципно различен експеримент.
На входа на усилвателя се подава слаб сигнал
$U_s(t)=U_s \cos(\omega t)$, който се усилва $Y$ пъти,
така че на $X$ входа на умножителя се подава напрежение
$U_X(t)=Y U_s(t)$.

При прерязана връзка между $X$ и $Y$ входовете на умножителя,
т.е. при отворен ключ S\textunderscore M от Фиг.~\ref{PCB},
към входа на платката означен с ``NC'' се подава известен сигнал 
$U_r(t)=U_r\cos(\omega t).$ Този сигнал отива до $Y$-входа на умножителя 
$U_Y(t)=U_s(t).$

Тогава умноженият сигнал съгласно Ур.~(\ref{eq:bmult}) ще бъде
\be
U_W=\frac12 \frac{YU_s(t) U_r(t) \cos^2(\omega t)}{U_\mathrm{m}}+U_Z.
\ee
Вече изразихме $U_Z$ чрез съпротивленията $R_1$ и $R_2$, затова директно използвайки  Ур.~(\ref{eq:buv}) и (\ref{eq:bmeas}), 
отчитайки и че $ \left < \cos^2(\omega t) \right > = 1/2$,
за усредненото по време напрежение, 
което мери волтметъра получаваме 
\be
U_V=\frac{Y U_r}{2 U_0}U_s .
\ee
Така на изхода на схемата волтметъра измерва постоянно напрежение $U_V$, в което участва и изследваният от нас периодичен сигнал с амплитуда $U_s$.
Сигналът $U_r$, който сме подали на Y входа се нарича носещ или референтен сигнал, а устройството чиято работа описахме току-що се нарича още локин (lock-in) волтметър (усилвател).

Локин усилвателите се използват за измерване на малки сигнали, 
дори когато на осцилоскоп 
усиленият сигнал се губи на фона на външния шум. 
При това синхронно измерване на два синусоидални сигнала с обща честота 
измервания сигнал може да бъде отделен от шума със спектрална плътност 
$(\mathcal{E}^2)_f$, ако
\be
U_s>\sqrt{(\mathcal{E}^2)_f \,\Delta f_\mathrm{av}},
\quad \Delta f_\mathrm{av}=\frac1{2\pi R_\mathrm{av}C_\mathrm{av}},
\ee
т.е. при достатъчно голямо време на усредняване 
$\tau_\mathrm{av}=R_\mathrm{av}C_\mathrm{av}$.
Все пак шумът не трябва да претоварва усилвателя
\be
U_m>\sqrt{Y^2\, (\mathcal{E}^2)_f\,B},
\quad B= \frac{f_0}{y_1}
\ee
и това поставя горна граница на възможния коефициент на усилване
\be
Y<\frac{U_m}{\sqrt{(\mathcal{E}^2)_f\,B}}.
\ee

\section{Отзиви за олимпиадата}

\begin{itemize}
\item Адаптация на компютърен превод от гръцки: 

Отзив за 5-та Балканска олимпиада по експериментална физика ``Ден на електронa''

Благодарение на участието в олимпиадата 
учителите получават ценна информация, 
запознават се как провежда това състезание и 
обменят мнения с колеги от други страни за образование и наука.
Бих искал да поздравя всички ученици от гръцкия отбор, участвали в конкурса, 
това е исторически първото гръцко участие.
Балканската олимпиада се организира от Съюза на физиците в България  
със съдействието на Физически факултет на Софийския университет Св. Климент Охридски.
През годините олимпиадата е създадена като традиция за извънкласно обучение по физика.
Целта на тазгодишната олимпиада беше да представи експериментален проблем, като изследва свойствата на електрона и флуктуациите на напрежението. 
Авторите на задачата са модифицирали експериментите, 
описани в ръководствата и учебници като част от учебната програма. 
В първия ден всички участници и техните придружители се събраха в аулата на Физически факултет. 
Това бяха 150 ученици от Балканите и Русия. 
Декана на Факултета (проф. Драйшу) прочете встъпителната реч,
после ръководителя на конкурса и от участници от различни страни. 
Ентусиазмът беше много голям.
След това учениците бяха разделени да работят в аудитории от около 15 души от различни националности. 
Те имаха четири часа за решаване на експерименталните задачи на олимпиадата , 
включващи експеримента предложен преди 111 години от великия физик Айнщайн -- иновативен метод за измерване на константата на Болцман чрез изучаване на средната топлинна енергия на кондензатор. 
Изненадващо, към днешна дата този експеримент не беше извършен,
така че групата от преподаватели от университета 
отчитайки развитието на методологията и технологиите през последните 100 години
са направили възможно да се адаптира този на научен проблем на училищно ниво.
И са първите, които го прилагат на практика
и го предлагат на участниците в конкурса.
Първоначално студентите получиха необходимите материали за извършване на експеримента, като електронни платки, батерии и проводници. 
Учениците трябваше да са запознати с използването на измервателните уреди (мултиметри), които те носеха със себе си. 
Докато учениците се състезаваха, учителите се събраха в друга аудитория и проведоха експеримента едновременно. 
Преди експеримента установено, че брошурите не бяха преведени на гръцки език, 
докато за участниците от другите страни условията бяха преведени на езика на който се преподава в съответната страна.
Гръцките участници използваха английския превод, 
което направи задачата малко по-трудна за пълното разбиране,
но въпреки това гръцките участници се представиха добре.
Организацията на конкурса беше безупречна.
На следващия ден се върнахме в Университета, 
където последва церемонията по връчване на наградите.
Най-добре се представи в конкурса един ученик от Казахстан. 
София ни посрещна с мечтателен снежен пейзаж.
\item 
Много беше забавно, искам пак!
\item 
Да има среща и беседа  с авторите на задачите. Може би след края на олимпиадата.
\item 
Задачите са повече като инструкции какво да се направи и не изискват голямо разбиране. Организацията и издаването на класация може да се подобрят, но с изключение на това идеята е страхотна и беше забавно. 
Надявам се в следващите години олимпиадата да се разпространи и утвърди. 
Благодаря ви, че я организирате за нас :) 
\item 
Участвам в олимпиадата с удоволствие.
Благодаря за положения огромен труд от ваша страна. 
Олимпиадата ми дава това, което училището не може.
\item 
Можеби да се зголеми котизацијата колку да се овозможат медали на најуспешните во сите категории, мислам дека дечките би биле презадоволни!!!
\item 
Задоволство ми беше да ги запознам моите колеги од другите држави. 
Тоа ме направи побогата и посигурна во моите сопствени вредности.
Задоволна сум што моите ученици од услови на потполна занемареност на експерименти во наставата по физика се одважија да се одмерат со своите врсници од другите држави. 
Верувам во нивната интелектуална вредност и неверојатната храброст искажана пред големите предизвици од овој вид.
Би сакала да има повеќе теми кои би им дале шанса на девојките да ги покажат своите способности. 
Толкаво изобилство од електроника некако ги обесхрабрува.
 Од толкав број на награди само една беше девојка или од оние што не беа присутни можеби имаше уште некоја.
Малку ми пречеше што немаше квалитетен интернет и еднаква можност за сите да учествуваат со полн капацитет во решвањето и испраќањето на решението. 
Мислам дека во тој дел не дојдовме целосно до израз како тим.
\item
Пожелно е да се случуваат некакви активности за менторите па и за учениците (на пример оние кои завршуваат порано) во текот на решавањето на задачата.
Би сакале да видеме поширока ранг листа, за да знаеме каков успех постигнале сите ученици.
\item 
Pristo sam zadovolen.
\item 
Учениците добро да бидат поделени во возрасни групи и сите да добиваат задача во зависност со нивната возраст. 
Најмалите од S категорија на пример да бидат наградувани и бодувани само за S категорија и да немаат право или да не се оценувани за M L и XL категорија.

\item 
The Olympiad was truly great and I had real fun conducting the experiment, which I plane to conduct with my classmates one day at school. I have a few recommendations for the next Olympiad as well. First of all, it would be good to be are informed about the instruments that we will have to use during the experiment. I personally had a problem with the potentiometer since this was the first time I ever got to connect it to the circuit and I had to spend time in order to find out how. This resulted in spending a big amount of time and then not having enough to design a nice graphical representation of the experimental data. Besides that, it would also be good to have at least one thermometer in every classroom in order to determine the room temperature, because guessing it makes the experiment inaccurate. Other than that, I believe the organization of the Olympiad was great. The experimental setup is truly amazing and I would like to say a big thank you for offering it to us for free. I will try to use it in the best way possible. It was an honour to have been part of this Olympiad.

\subsection{Отговори на въпроса: Повторихте ли експеримента в къщи? Каква стойност за константата на Болцман получихте?}

\item Да  повторихме го в час по физика.
\item Да, 15\% грешка.
\item $1.29\times 10^{-23}$J/K.
\item I have not repeated the experiment at home yet, but I am planning to do so in a week during the Christmas holidays, since I am going to have a lot of free time. Nevertheless, I have actually kept the measurements I did during the EPO and I analyzed them using Microsoft Office Excel. The value I got was k=1.182*10$^{-23}$~ J/K  (at a temperature of 17 degrees Celsius) which is not very close ( there is a 15\% error compared to the known value, k=1.38*10$^{-23}$~J/K). I concluded that I should have measured the voltage U2,n ( the one measured in task 18) more precisely, in mV.   

\end{itemize}

\section{Резултатът на абсолютният шампион}

\begin{figure}[h]
\includegraphics[scale=0.36]{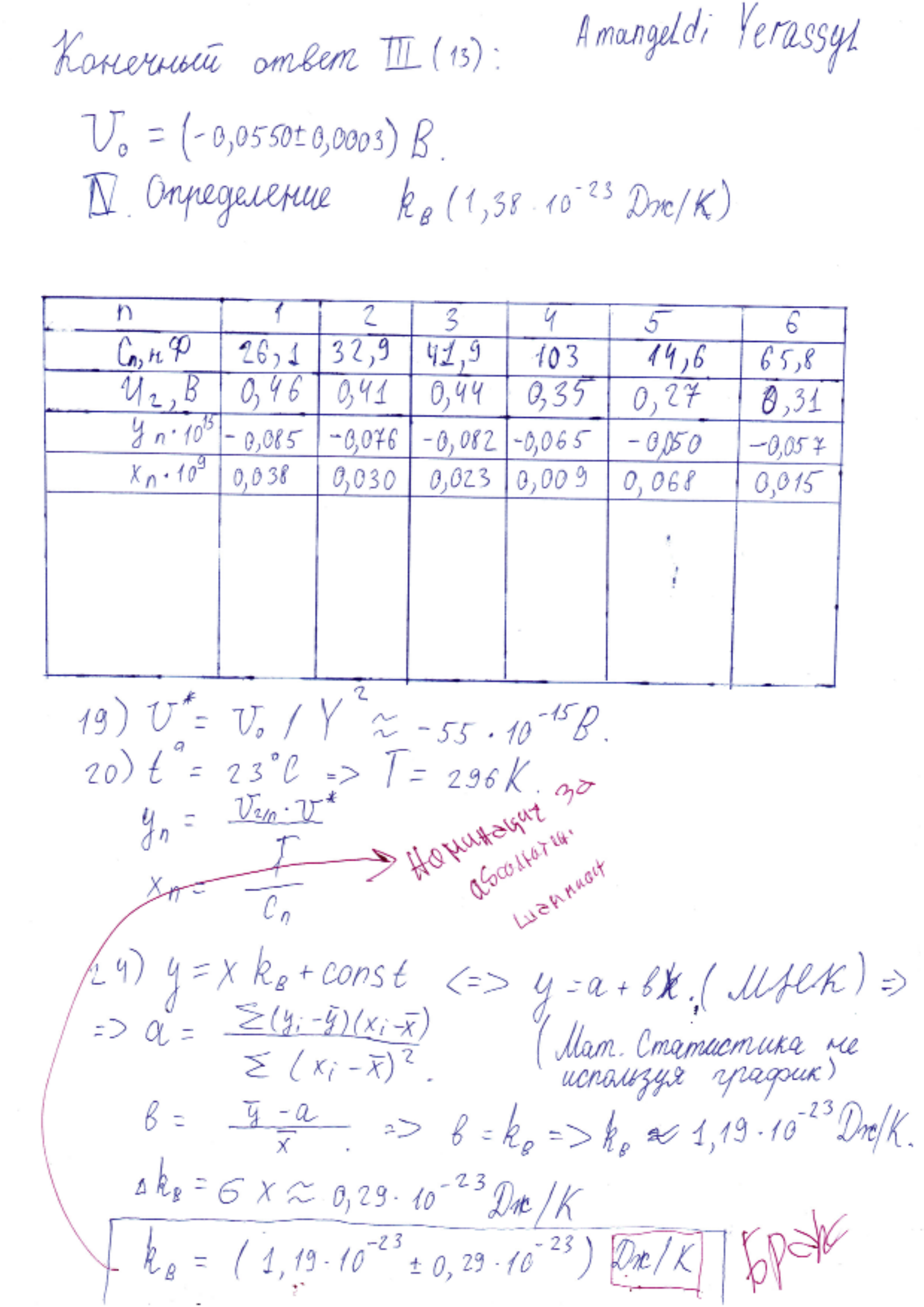}
\end{figure}

\end{document}